\documentclass[twocolumn]{aastex63}
\usepackage{apjfonts, natbib, lineno, hyperref, url}
\usepackage{longtable, booktabs, array, amsmath}

\newcommand{\ha}{H\ensuremath{\alpha}}
\newcommand{\hb}{H\ensuremath{\beta}}
\newcommand{\nii}{[N\,{\footnotesize II}]}
\newcommand{\sii}{[S\,{\footnotesize II}]}
\newcommand{\mgii}{Mg\,{\footnotesize II}}
\newcommand{\hei}{He\,{\footnotesize I}}

\newcommand{\oiii}{[O\,{\footnotesize III}]}

\newcommand{\lum}{erg~s\ensuremath{^{-1}}}
\newcommand{\dd}{$\Delta d$}
\newcommand{\wmax}{\mathrm{W2_m}}
\newcommand{\lbol}{\ensuremath{L\mathrm{_{bol}}}}

\newcommand{\lw}{log\ensuremath{L\mathrm{_{W2}}}}
\newcommand{\ms}{log\ensuremath{M\mathrm{*}}}
\newcommand{\msun}{\ensuremath{M_{\odot}}}
\newcommand{\mb}{\ensuremath{M_\mathrm{bulge}}}
\newcommand{\kms}{\ensuremath{\mathrm{km~s^{-1}}}}
\newcommand{\mbh}{\ensuremath{M_\mathrm{BH}}}
\newcommand{\wise}{\emph{WISE}}
\newcommand{\neowise}{\emph{NEOWISE}}
\newcommand{\spitzer}{\emph{Spitzer}}

\newcommand{\ldust}{\ensuremath{L_\mathrm{dust}}}

\newcommand{\mstar}{\ensuremath{M\mathrm{_{\star}}}}
\newcommand{\vdisp}{\ensuremath{\sigma\mathrm{_{\star}}}}

\shorttitle{MIRONG Sample}
\shortauthors{Jiang et al.}

\begin{document}
%\linenumbers

\title{Mid-InfraRed Outburst in Nearby Galaxies (MIRONG) I: Sample Selection and Characterization}
\correspondingauthor{Ning Jiang}
\email{jnac@ustc.edu.cn}

\author[0000-0002-7152-3621]{Ning Jiang}
\affiliation{CAS Key laboratory for Research in Galaxies and Cosmology,
Department of Astronomy, University of Science and Technology of China, 
Hefei, 230026, China; jnac@ustc.edu.cn}
\affiliation{School of Astronomy and Space Sciences, 
University of Science and Technology of China, Hefei, 230026, China; twang@ustc.edu.cn}

\author[0000-0002-1517-6792]{Tinggui Wang}
\affiliation{CAS Key laboratory for Research in Galaxies and Cosmology,
Department of Astronomy, University of Science and Technology of China, 
Hefei, 230026, China; jnac@ustc.edu.cn}
\affiliation{School of Astronomy and Space Sciences,
University of Science and Technology of China, Hefei, 230026, China; twang@ustc.edu.cn}

\author[0000-0002-4757-8622]{Liming Dou}
\affiliation{Department of Astronomy, Guangzhou University, Guangzhou 510006, China;
doulm@gzhu.edu.cn}

\author[0000-0002-7020-4290]{Xinwen~Shu}
\affiliation{Department of Physics, Anhui Normal University, Wuhu, Anhui, 241000, People's Republic of China}

\author{Xueyang~Hu}
\affiliation{CAS Key laboratory for Research in Galaxies and Cosmology,
Department of Astronomy, University of Science and Technology of China, 
Hefei, 230026, China; jnac@ustc.edu.cn}
\affiliation{School of Astronomy and Space Sciences,
University of Science and Technology of China, Hefei, 230026, China; twang@ustc.edu.cn}

\author{Hui~Liu}
\affiliation{CAS Key laboratory for Research in Galaxies and Cosmology,
Department of Astronomy, University of Science and Technology of China, 
Hefei, 230026, China; jnac@ustc.edu.cn}
\affiliation{School of Astronomy and Space Sciences,
University of Science and Technology of China, Hefei, 230026, China; twang@ustc.edu.cn}

\author{Yibo~Wang}
\affiliation{CAS Key laboratory for Research in Galaxies and Cosmology,
Department of Astronomy, University of Science and Technology of China, 
Hefei, 230026, China; jnac@ustc.edu.cn}
\affiliation{School of Astronomy and Space Sciences,
University of Science and Technology of China, Hefei, 230026, China; twang@ustc.edu.cn}

\author[0000-0003-1710-9339]{Lin~Yan}
\affiliation{
Caltech Optical Observatories, California Institute of Technology, Pasadena, CA 91125, USA
}

\author[0000-0001-6938-8670]{Zhenfeng~Sheng}
\affiliation{CAS Key laboratory for Research in Galaxies and Cosmology,
Department of Astronomy, University of Science and Technology of China, 
Hefei, 230026, China; jnac@ustc.edu.cn}
\affiliation{School of Astronomy and Space Sciences,
University of Science and Technology of China, Hefei, 230026, China; twang@ustc.edu.cn}

\author[0000-0003-4975-2333]{Chenwei~Yang}
\affiliation{Polar Research Institute of China, 451 Jinqiao Road, Shanghai, 200136, People's Republic of China}

\author[0000-0002-7223-5840]{Luming~Sun}
\affiliation{CAS Key laboratory for Research in Galaxies and Cosmology,
Department of Astronomy, University of Science and Technology of China, 
Hefei, 230026, China; jnac@ustc.edu.cn}
\affiliation{School of Astronomy and Space Sciences,
University of Science and Technology of China, Hefei, 230026, China; twang@ustc.edu.cn}
\affiliation{Department of Physics, Anhui Normal University, Wuhu, Anhui, 241000, People's Republic of China}

\author[0000-0003-1956-9021]{Hongyan~Zhou}
\affiliation{CAS Key laboratory for Research in Galaxies and Cosmology,
Department of Astronomy, University of Science and Technology of China,
Hefei, 230026, China; jnac@ustc.edu.cn}
\affiliation{Polar Research Institute of China, 451 Jinqiao Road, Shanghai, 200136, People's Republic of China}

\begin{abstract}
The optical time-domain astronomy has grown rapidly in the past decade 
but the dynamic infrared sky is rarely explored.
Aiming to construct a sample of mid-infrared outburst in nearby galaxies (MIRONG),
we have conducted a systematical search of low-redshift ($z<0.35$) SDSS 
spectroscopic galaxies that have experienced recent MIR flares using
their \emph{Wide-field Infrared Survey Explorer} (\wise) light curves. 
A total of 137 galaxies have been selected by requiring a brightening 
amplitude of 0.5 magnitude in at least one \wise\ band with respect to 
their quiescent phases.	
Only a small faction (10.9\%) has corresponding optical flares.
Except for the four supernova (SNe) in our sample, the MIR luminosity of 
remaining sources ($L_{\rm 4.6\mu m}>10^{42}$~\lum) are markedly brighter 
than known SNe and their physical locations are very close 
to the galactic center (median $<0\arcsec.1$). 
Only four galaxies are radio-loud indicating synchrotron radiation 
from relativistic jets could contribute MIR variability. 
We propose that these MIR outburst are dominated by the dust echoes of transient
accretion onto supermassive black holes, such as tidal disruption events (TDEs) 
and turn-on (changing-look) AGNs.
Moreover, the inferred peak MIR luminosity function is generally consistent with 
the X-ray and optical TDEs at high end albeit with large uncertainties.
Our results suggest that a large population of transients have been overlooked by
optical surveys, probably due to dust obscuration or 
intrinsically optical weakness . 
Thus, a search in the infrared band is crucial
for us to obtain a panoramic picture of nuclear outburst.
The multiwavength follow-up observations of the MIRONG sample 
are in progress and will be presented in a series of subsequent papers.
\end{abstract}


\section{Introduction}
\label{intro}

Time-domain astronomy has developed rapidly in the past two decades.
%The time-domain astronomy rapidly developed in the past two decades 
%have brought about the flourishing study of extragalactic sky 
%in a dynamic way instead of merely static. 
The great progress has been driven by the advent of new instruments and 
facilities dedicated to wide-field, deep and fast surveys, such as  
Catalina Real-Time Survey (CRTS; \citealt{Drake2009}), 
Palomar Transient Factory (PTF/iPTF; \citealt{Law2009}), 
Panoramic Survey Telescope and Rapid Response System (Pan-STARRS or PS; 
\citealt{Kaiser2004}; \citealt{Chambers2016}), 
All Sky Automated Survey for SuperNovae (ASASSN; \citealt{Shappee2014}), 
the Asteroid Terrestrial-impact Last Alert System (ATLAS; \citealt{Tonry2018})
and Zwicky Transient Facility (ZTF; \citealt{Graham2019}).
These remarkable projects have gradually make it possible to image 
the entire sky every a few days and process the data in real time.
It is believed that 
%the dynamic extragalactic science will be a
%fully mature field in the 2020s with the arrival of LSST, which is an
%unprecedented edge tool of time-domain survey.
LSST will produce an unprecedented time-domain survey in the 2020s.

Supernova (SNe) are absolutely the major class of extragalactic transients 
and its number has experienced explosive growth as expected 
with aid of those above-mentioned surveys.
In 2018, there were more than 7,000 reported supernovae
\footnote{http://rochesterastronomy.org/sn2018/index.html}
although most have not been spectroscopically confirmed.
In addition to increasing in number of SNe, these surveys
have also accelerated the discovery 
and in-depth studies of peculiar SNe, such as superluminous supernova 
(SLSNe; e.g., ASASSN-15lh, \citealt{Dong2016}; 
see \citealt{Gal-Yam2019} as a recent review), 
exotic SNe with multiple peaks (e.g., iPTF14hls, \citealt{Arcavi2017})
and even gravitationally lensed SNe (iPTF16geu, \citealt{Goobar2017}).
These unusual events could make great breakthrough in our understanding 
of stellar explosion in extreme physical condition.

The other population of extragalactic variable sources which has aroused
great attention is the transient events associated with 
the supermassive black holes (SMBHs) located in the centers of galaxies. 
Among them, the tidal disruption event (TDE), in which a star is 
torn apart by a SMBH's the tidal force, is of particular interest.
During the process, about half of the stellar mass may be ejected while 
the rest of the stellar material is accreted on to the black hole, 
producing a luminous flare of electromagnetic radiation lasting for months to 
years (\citealt{Rees1988,Evans1989,Phinney1989}).
TDE discoveries require modern time-domain surveys
because their event rate is hundreds of times lower than SNe with a rate 
of $10^{-4}-10^{-5}$/galaxy/year (\citealt{Wang2004,Stone2016}).
Hence, the number of TDEs  (or candidates) found to date is still very limited.
%only a few of dozens\footnote{https://tde.space/}. 
As a rare and special form of accretion, TDEs are nevertheless extremely 
scientifically valuable as they offer us an ideal chance to 
probe the existence, mass and spin of SMBHs in normal galaxies 
(\citealt{Lu2017,Mockler2019,Pasham2019}).
Moreover, they can serve as a unique laboratory to study the dynamic process of 
BH activity by witnessing the ignition and flameout of the accretion disk
(e.g., \citealt{Wevers2019}) as well as rapidly-launched jets 
(e.g., \citealt{Bloom2011,Burrows2011,Mattila2018}).

Despite occasionally swallowing stars, SMBHs are believed to grow mainly by 
accreting surrounding gas during their active galactic nuclei (AGN) phases.
Stochastic variability is ubiquitous in AGNs, among which a small fraction 
are extremely variable with light curve featured as flaring 
(e.g., \citealt{Graham2017}) or state changing (e.g., \citealt{Graham2020}) pattern.
Follow-up spectroscopic observations of the most variable AGNs suggest
they can even change their types on time scales of years, 
characterized by appearance or disappearance of broad emission lines 
(e.g., \citealt{Shappee2014,LaMassa2015,Runnoe2016,MacLeod2016,Yang2018}). 
These objects are dubbed as changing-look (CL) AGNs, and their variation 
is driven by the dramatic change of the accretion flow rather than the obscuration
(\citealt{Sheng2017,Hutsemekers2019}).
The most intriguing subclass could be the so-called "turn-on" AGNs, 
which transition from a quiescent galaxy to a type 1 AGN within several months to years. 
Although only few such systems have been confirmed (\citealt{Gezari2017,Yan2019}; see also \citealt{Frederick2019}), 
they appear to not be extremely rare, and even one
poses challenges to canonical accretion disk theories.

In a nutshell, the known extragalactic transient sky is dominated by SNe and 
transient SMBH accretion. 
The investigation of latter has been facilitated recently by improved time-domain surveys.
Current large surveys have been exclusively performed in the optical band,
which are blind to transients that are either self-obscured (e.g. type~2 AGNs) 
or located in the dusty regions (e.g. SNe).
Therefore, it is crucial to conduct surveys in bands free of dust obscuration.
In this sense, the mid-infrared (MIR) is the most promising band but it 
was completely unexplored until the \emph{spitzer} Deep Wide-Field Survey, which
has now been incorporated into the Decadal IRAC Bo{\"o}tes Survey. This survey
is used to search for obscured SNe and to study the quasar variability in the MIR
\citep{Kozlowski2010,Kozlowski2016}
by taking advantage of the repeatedly surveyed region ($\sim9~\rm deg^2$) .
SPitzer InfraRed Intensive Transients Survey (SPIRITS, \citealt{Kasliwal2017}) 
is a more recent project dedicated to finding infrared luminous transients 
by targeting 190 nearby galaxies, which has yielded out numerous hidden 
SNe and dusty stellar outburst (\citealt{Jencson2019}).
Despite this, such projects generally focused on a small sample 
of galaxies or a small sky region, so they are almost incapable 
of capturing rare events like TDEs as well as "turn-on" AGNs.

Fortunately, the \emph{Wide-field Infrared Survey Explorer} (\wise) 
mission (\citealt{Wright2010}) and its asteroid-characterization extension, 
the Near-Earth Object Wide-field Infrared Survey Explorer Reactivation (\neowise) 
mission (\citealt{Mainzer2014}), have opened a new window to map the MIR
dynamic sky benefited from its ideal survey mode (see \S\ref{wisemode}).
For instance, \citet{Sheng2020} have come up with an efficient method to 
look for CL AGNs from MIR-variable quasars screened by 
\wise\ light curves (see also \citealt{Stern2018,Assef2018}) but 
with the caveat that only "turn-off" CL AGNs can be selected because of 
their parent sample of quasars.
\citet{Wang2018} used \emph{WISE} to systematically search for TDEs in
extremely variable normal galaxies and discovered IR echoes of TDEs,
immediately after the pioneering discoveries of IR echoes of TDEs 
(\citealt{Jiang2016,Dou2016,vV2016}).
However, \citet{Wang2018} select variables using the variability flag
given by \wise\ pipeline, which is on basis of data taken 
between 2009 December and 2011 February.
As a result, the acquired galaxies have all entered the stage of dimming, 
which make the further confirmation of their physical nature unrealistic. 

To overcome the limitation from \citet{Wang2018}, 
we have designed a new project to search for and explore
MIR Outburst in Nearby Galaxies (MIRONG) specifically.
MIRONG uses both the public \wise\ and \neowise\ databases to search 
for more recent outburst events in galaxies and enable multi-wavelength follow-up. 
MIRONG may uncover a population of extragalactic transients which 
have been overlooked by traditional optical surveys and ultimately
improve our understanding of the extragalactic dynamic sky.
The paper is organized as follows. 
In Section 2, we describe the process of sample selection of MIRONG. 
In Section 3-5, we characterize the sample from 
properties of MIR light curves, event rate and luminosity function, and
their host galaxies, respectively. 
We inspect the nature of MIRONG in Section 6 and discuss the 
implications of our work in Section 7.
Finally, we end with a brief summary and discussion of future prospects in Section 8. 
We assume a cosmology with $H_{0} =70$ km~s$^{-1}$~Mpc$^{-1}$, $\Omega_{m} = 
0.3$, and $\Omega_{\Lambda} = 0.7$.

\section{Sample selection}
\label{sample_selection}

\subsection{Characteristics of WISE and NEOWISE Survey}
\label{wisemode}

\wise\ has performed a full-sky imaging survey in four broad bandpasses 
centered at 3.4, 4.6, 12 and 22~$\mu$m (labeled W1-W4) from 2010 January to August.
\wise\ continued surveying the sky in its bluest three
bands during 2010 August and September. After that, its \neowise\ 
hunted asteroids until 2011 February (\citealt{Mainzer2011}), 
with only the W1 and W2 channels remaining operational
as the solid hydrogen cryogen used to cool the W3 and W4 instrumentation 
has been depleted. 
Following a 33 month hibernation period, the \wise\ instrument recommenced survey
operations in 2013 December (\citealt{Mainzer2014}).
This post-hibernation mission is referred to as NEOWISE-Reactivation  
(\neowise-R) to hunt for asteroids that could pose an impact hazard to the Earth.
 
The \wise\ survey strategy is very novel. 
It has a field-of-view of $47^\prime\times47^\prime$ and a small (10\%) 
overlap between adjacent fields in one orbit. The scan circle advances 
by about 4$^\prime$ per orbit and 15 orbits can be fulfilled each day.
Therefor, for a typical sky region, the available exposures can be 
segmented into a series of 1 day time intervals (referred to as visits), 
with such visits occurring once every six months except for the gap
during the hibernation.
As a spontaneous outcome of the unique survey mode,
it becomes an unprecedented database to study the transient MIR sky.
There are typically 12 successive orbits covering a given source within 
one day, with denser coverage toward higher ecliptic latitude 
(\citealt{Wright2010,Hoffman2012}).
Such a high-frequency sampling allows us to probe the intraday MIR variability, 
which was seldom explored in the past 
(e.g., \citealt{Jiang2012,Jiang2018}).

The \neowise\ survey is still ongoing at the time of submission of this paper.
Our sample selection is based on the \wise\ and \neowise-R data 
from 2010 to the end of 2018, that is all of the available data when 
we did this work.  
They have yielded an average of 12 visits for each target,
so an investigation of MIR variability on year time scales
is absolutely achievable for a large sample as proved by previous works
(e.g., \citealt{Wang2018}).
We have noticed that the latest \neowise\ data release (2020 March 26)
has come out just before our paper submission, 
while it is a massive work to redo our work from the very beginning by including
the newly released date from 2018 December to 2019 December.
However, the newest photometry has been added in the analysis of 
our final sample of MIRONG since the more complete light curves 
will help us obtain more accurate quantities (e.g., peak luminosity).

\subsection{Parent Sample}

The Sloan Digital Sky Survery (SDSS) has spectroscopically observed 
millions of galaxies with apparent Petrosian magnitude $r<17.77$ 
(\citealt{Strauss2002}).
We have checked the SDSS DR14 spectroscopic catalog (\citealt{Abolfathi2018}) 
and picked out those flagged with "GALAXY" class at $z<0.35$, that is 1,253,962 
spectra in total. Note that some of them are repeated multi-epoch data 
for same objects, there are 1,150,901 unique objects 
differentiated by their celestial coordinates.
We choose spectroscopic galaxies instead of photometric because 
we need to know our galaxies' redshift (z<0.35) and other properties.
The redshift cut is to ensure that the \ha\ region is located in the
wavelength range of SDSS spectrum, which will help us diagnose the spectral type
by emission-line ratios in the Baldwin-Phillips-Terlevich diagnostic diagram
(BPT diagram in short, \citealt{Baldwin1981}).
Second, the cut of z<0.35 will keep the dust emission within the \wise\ bands
as it is expected to start at 2~$\mu$m.
Furthermore, the signal-to-noise ratio (S/N) of more distant galaxies are 
generally too low to undertake variability studies accurately.
We designate these $\sim1$ million galaxies the parent sample.

\begin{figure*}
\epsscale{0.75}
\plotone{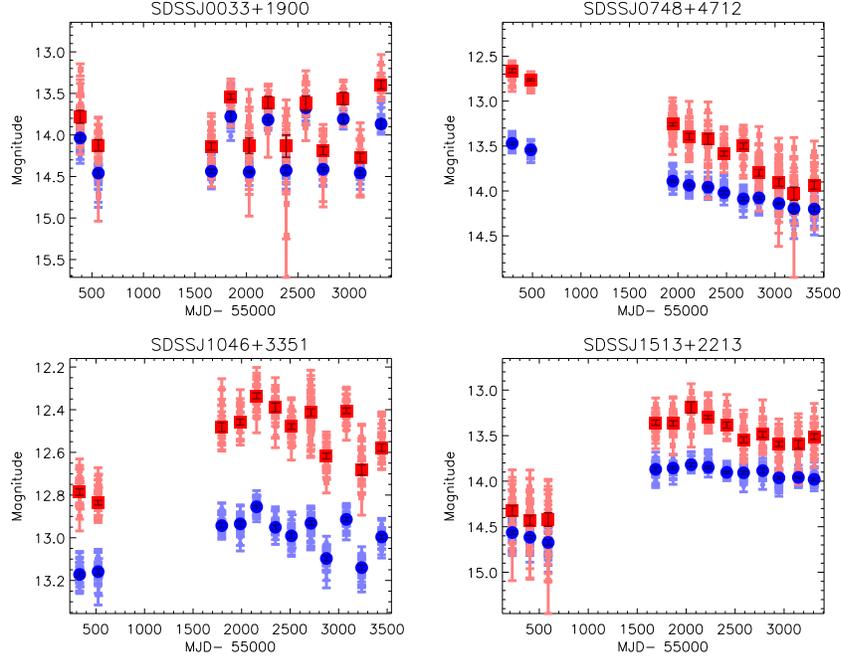}
\caption{
We show the different \wise\ light curves patterns of the MIR-variable galaxies, 
including the periodic oscillation (top left), long-term declination (top right), 
stochastic variations (bottom left) and state transition (bottom right).
The recent flare-like patterns are presented separately in Figure~\ref{lc}.
Blue dots: W1 ($3.4\mu$m); red squares: W2 ($4.6\mu$m). 
The raw single exposures are plotted in light blue and red, while the binned
data are plotted in dark blue and red.
As we explained in \S\ref{mirvar}, the case of one-year cycle 
oscillation (top left) is not physical but caused by the 
latent image artifact from a nearby bright star.
}
\label{pattern}
\end{figure*}

\subsection{MIR-variable Galaxies}
\label{mirvar}

We construct a sample of MIR-variable galaxies in
advance of the final selection of outbursts.
We retrieved the W1 and W2 profile-fit photometry of each galaxy 
in the parent sample from the public AllWISE Multiepoch Photometry Table and 
\neowise-R Single Exposure (L1b) Source Table\footnote{
https://irsa.ipac.caltech.edu/cgi-bin/Gator/nph-scan?mission=irsa\&submit=Select\&projshort=WISE},
encompassing all exposures from 2010 up to 2018.
The photometry is measured by point spread function (PSF) profile fitting,
in which the PSFs have been estimated from
observations of many tens of thousands of stars.
The AllWISE Multiframe pipeline detects sources on the deep coadded atlas images, 
and then measures the sources for all available single-exposure images 
in all bands simultaneously while the \neowise\ magnitudes are obtained by PSF-fit 
to individual exposures directly.

The acquired single-exposure data are first filtered by the quality flags marked 
in the catalogs. The bad data points with poor quality frame (\texttt{qi\_fact}<1),
charged particle hits (\texttt{saa\_sep}<5), scattered moon light (\texttt{moon\_masked}=1)
and artifacts (\texttt{cc\_flags}$\neq$0) have been removed.
In addition, we have also abandoned the photometry fitted with multiple PSF 
components (\texttt{nb}>1 and \texttt{na}>0), that is performed when the source is fit  
concurrently with other nearby detections or when a single object is split into 
two components during the fitting process.
The surviving data are immediately binned every half year to increase 
the S/N of the photometry, resulting in average 13 epochs for each target.
This binning strategy fits to the sampling rate of \wise\ survey (see \S\ref{wisemode}).
We begin to perform a blind search of variable galaxies which satisfy 
$\delta$W1$>$0.5 or $\delta$W2$>$0.5. 
$\delta$W1 ($\delta$W2) are the difference between the maximum and minimum values 
of W1 and W2 across the combined  ALLWISE+\neowise\ light curves.
%The 0.5 mag difference is applied given our ultimate goal of building
%a sample of huge outburst with brightened amplitudes larger than this value.
We have further requested $\delta$W1$>$0.3 or $\delta$W2$>$0.3 
during the \neowise\ phase to make sure the variability is 
still obvious after 2013 December for ease of follow-up studies. 
The latter condition has also excluded fake variable sources 
caused by possible systematical offset between ALLWISE and \neowise\ photometry.
The significance of the variability are required to be not lower than 
5$\sigma$, that is 
$(W?max-W?min)/\sqrt{(W?max\_{err})^2+(W?min\_{err})^2}>5$, 
to ensure that the variability is still valid when taking uncertainties 
into consideration.
Last, we have casted away faint sources by the cut 
$W2min<14$ to obtain a magnitude limit sample.

\begin{figure*}
\epsscale{0.75}
\plotone{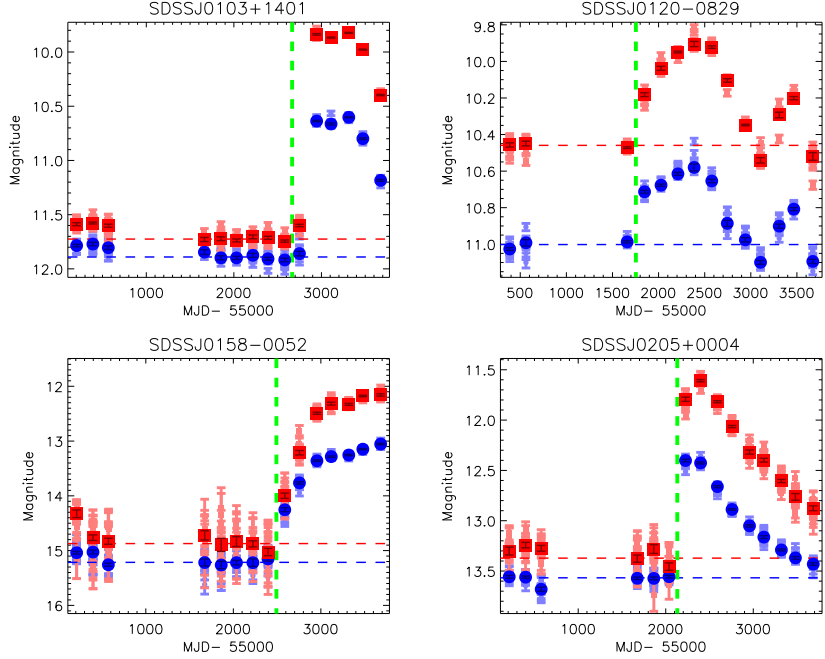}
\caption{
Similar to Figure~\ref{pattern}, the \wise\ light curves of four representative 
outburst galaxies in our MIRONG sample. The estimated magnitudes at the 
quiescent phase are shown with dashed blue and red lines. 
The vertical green dashed lines mark the boundary of the quiescent 
and outburst stage.
SDSSJ0103+1401 shows very fast rising followed by a relatively stable high state, 
in contrast with SDSSJ0205+0004, which displays also fast rising but immediate 
declining. The other two objects shows a slower rising, in which SDSSJ0120-0829
has manifested itself as a complete flare in terms of both rising and declining 
(\citealt{Sun2020}). 
}
\label{lc}
\end{figure*}

The above cut of variability amplitude has resulted in a sample consisting of 
1,026 galaxies\footnote{The general information of the 1,026 variable galaxies
can be downloaded from this website: 
\url{http://staff.ustc.edu.cn/~jnac/data\_public/wisevar.txt}}, 
allowing for us to check visually their MIR light curves one by one.
%This sample can be treated as a valuable collection
%of MIR-variable galaxies with diverse light curve patterns.
While stochastic variability is predominant, 
peculiar variability patterns are also visible (see Figure~\ref{pattern}), 
such as state transition, long-term declining and flare-like rising light curves. 
We notice that very few of them ($11$ objects) show periodic oscillations 
with a cycle of one year (see the example in the top left panel of Figure~\ref{pattern}).
These objects are affected by the latent image artifact from a nearby bright star 
that appears in the preceding image in the scan\footnote{
The \wise\ scan direction on the sky flips every 6 months as the orbit 
precesses around the sky. Hence a particular point on the sky will 
alternate being scanned north-to-south and south-to-north each 6 months. 
This means that the bright star that is seen on the preceding image in one epoch 
will be seen on the following image 6 months later. 
As a consequence, the artifact affects with one year cycle.
The polluted photometry should have been flagged as "P" in
"cc\_flags" but the flag does not work well when the source is slightly
offset from the predicted position of the latent and
the expected size of the latent is poorly modeled.
}.
The photometry pipeline has often failed to flag them automatically
%as it is still imperfect for this sort of pollution.
%In brief, 
so that the one-year cycle variations may not be real.
In addition, we emphasize that our selection of the MIR variable 
sources is conservative, % according to the cut of variability amplitude, 
% at the cut of variability amplitude and significance
%in view of our strict demand on the variability amplitude and significance.
and the actual fraction of these galaxies may be much higher than $0.1\%$ 
as we inferred here.

\subsection{Sample of MIRONG: Galaxies with MIR Outburst}

\begin{figure}
\epsscale{1.1}
\plotone{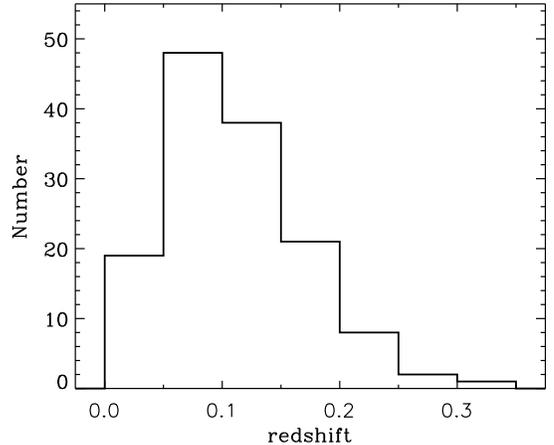}
\caption{
The histogram of redshift of our sample.
}
\label{zdis}
\end{figure}

%Focusing on our main objective, the MIR outburst galaxies are undoubtedly a 
%subsample of the variable ensemble above.
%Specifically, we aim to pick out the well-defined outbursts, 
For the purpose of this study, we further selected a subsample 
of MIR outburst galaxies, which initially display a stable phase yet followed by a significant 
brightening seen in the MIR light curves.
The emission in the quiescent state is served as the background that will be subtracted from the 
outburst light curves.
We estimated the pre-outburst magnitudes
by adopting the median value of the data points in the quiescent state. 
The beginning of outburst is set to the epoch when either W1 or W2 magnitude
shows a brightening over $3\sigma$ significance.
We adopted two criteria to qualify a flare: 
(1) the maximum flux density after outburst has brightened by $>0.5$ mag in W1 or W2 with respect to the quiescent state ($\Delta$W1$>0.5$ or $\Delta$W2$>0.5$); 
(2) the variability significance at the epoch of maximum flux 
is larger than 5$\sigma$ in at least one band.
The above criteria are effective, with which we yield 148 MIR outburst sources.
Lastly, we caution that the \wise\ spatial resolutions at W1 and W2 bands are of 
$\sim6\arcsec$, so the photometry could suffer from the contamination of nearby sources. 
Thus, we rejected those targets with companions (galaxy pairs or polluted by 
foreground stars) within project distance closer than 6\arcsec\ 
by checking their SDSS images visually.
As a result, 137 objects are left, constituting the final outburst 
sample (see basic parameters in Table~\ref{tb1_sample} and 
four representative examples in Figure~\ref{lc}).
The median redshift is 0.102 (see Figure~\ref{zdis}).

To summarize, we have finally selected a well-defined sample of 137 MIR outburst 
nearby galaxies (MIRONG) which display a brightening over 0.5 mag with respect to 
the previous quiescent phase in the \wise\ light curves. 
These galaxies are all initially drawn from the SDSS spectroscopic catalog 
with rich available information, e.g., properties 
of host galaxies and nuclear activity (see \S\ref{host}), 
which allows for further detailed study on the nature of transient 
MIR emission in the center of galaxies. 

%They are excellent representatives of the most dramatic MIR transients associated 
%with galaxies and are worthwhile to study in depth in virtue of their rich archival 
%data and by proposing follow-up observations.

\section{Properties of MIRONG}

\subsection{Optical Counterparts}
\label{opt}

It is natural to ask  whether these MIR flares have been detected by 
various optical surveys.
First, we cross-matched our sample with (candidate) 
SNe catalog (\citealt{GalYam2013})\footnote{http://rochesterastronomy.org/supernova.html, 
noting that they are not all real SNe, but candidates including nuclear transients.}
discovered between 2010 and 2018 and yielded 11 objects. 
In addition, we have also added four more objects that are reported in literature
but not listed in the SNe catalog, including two Gaia nuclear 
transients (SDSSJ0854+1113, SDSSJ1647+3843; \citealt{Kos2018}),
one CRTS outburst event (SDSSJ1332+2036, \citealt{Drake2019}) and another 
turn-on quasar discovered by iPTF (SDSSJ1554+3629, \citealt{Gezari2017}). 
The basic information about the known 15 optical transients is presented 
in Table~\ref{optical_match}.

Among these sources, 
%a few of them are well-studied and classified cases.
SDSSJ0936+0615, SDSSJ1531+3724, SDSSJ1540+0054 and SDSSJ1554+1636 are 
spectroscopically confirmed SNe discovered by ASAS-SN.
SDSSJ0158-0052 is a candidate TDE discovered by Pan-STARRS and ASAS-SN, 
which occurred in a Seyfert 1 galaxy with low mass black hole 
(\citealt{Blanchard2017}), whose MIR flare has been successfully
explained by the dust echo of the AGN torus (\citealt{Jiang2017}). 
SDSSJ1554+3629, SDSSJ0915+4814 and SDSSJ1133+6701 are newly reported turn-on AGNs 
which have transferred from LINERs to quasars within a few of years 
(\citealt{Gezari2017}; \citealt{Frederick2019}).
SDSSJ1620+2407 is a candidate TDE discovered by ATLAS, which shows markedly 
newly-appeared broad Balmer lines and blue continuum (\citealt{Fraser2017}).
Apart from the nine studied objects above, the remaining six sources have only been alerted 
by photometry without further spectroscopic observations.
Despite lack of reliable identifications, in contrast with the four SNe cases (>1\arcsec), 
their close distance to the galactic center (<1\arcsec) suggests  
that they are likely associated with the SMBH activity rather than stellar explosion.

For the remaining 120 galaxies for which no known optical counterparts 
are matched, we have also examined the public 
%CRTS\footnote{http://nunuku.caltech.edu/cgi-bin/getcssconedb\_release\_img.cgi} 
%and ASAS-SN\footnote{https://asas-sn.osu.edu/} photometric data. 
CRTS (\citealt{Drake2009}) and ASAS-SN (\citealt{Kochanek2017}) data.
The combined data give $V$-band light curves spanning from 2005 to 2018.
We did not find any optical flares, implying the faintness of optical 
emission for most, if not all, MIR bursts galaxies.

\begin{figure}
\epsscale{1.1}
\plotone{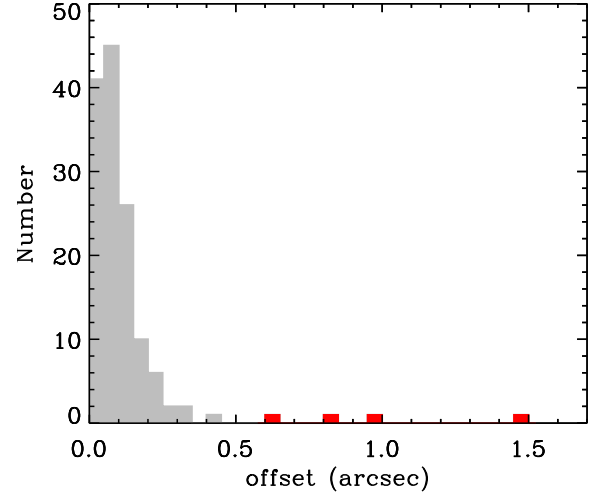}
\caption{
The histogram of center offset distance (\dd) in unit of arcsec, 
defined as the angular distance of the photometric center at the maximum epoch 
and the quiescent epoch.  
The \dd\ of SNe (highlighted in red) are all larger than 0\arcsec.5 in 
contrast with the tiny values of other objects (median 0\arcsec.08). 
}
\label{offcenter}
\end{figure}

\subsection{Constrains on the Physical Position of MIR flares}

The specific locations in the galactic region of the MIR flares 
can give us strong clues to their nature. For example, an outburst 
offset from the galactic center can be convincingly classified as SNe.
Otherwise the central SMBHs might be considered as the most likely origin. 
%albeit the SNe close to the nucleus can not be completely ruled out.

The astrometry of \wise\ catalog has been reconstructed with respect to the 2MASS 
Point Source Catalog reference frame. The root mean square error of the position 
is found to be less than 0.5\arcsec, for sources with SNR$>$20 in at least 
one \wise\ band, where the noise includes flux errors due to 
zodiacal foreground emission, instrumental effects, 
source photon statistics, and neighboring sources\footnote
{http://wise2.ipac.caltech.edu/docs/release/allsky/expsup/sec6\_4.html}.
The \wise\ PSF profile-fit photometry we used is measured at the 
intensity-weighted center, so the variation of the center position can tell us 
whether or not the outburst is far away from the pre-outburst galactic center. 
For instance, when an off-centered SNe has exploded and is 
MIR bright, there should be a positional offset between the \wise\ photometric center 
of its host galaxy and the intensity-weighted center determined by the SNe outburst.
%the accompanied \wise\ photometric center of its host galaxy
%should be shifted with an accurate distance determined by the SNe MIR luminosity.

To test whether there significant positional offsets exist, we have calculated 
the mean photometric center (RA, DEC) given by ALLWISE and \neowise-R catalog
at the quiescent state and that from the maximum outburst epoch, respectively.
We defined the angular distance between them as the offset distance (\dd, in unit of arcsec).
The distribution of \dd\ is presented in Figure~\ref{offcenter}.
The majority of our galaxies show very small \dd\ with a median
value of $0\arcsec.08$. 
We noticed that only the four known SNe (see \S\ref{opt}) have \dd\ 
larger than $0\arcsec.5$ (1.48, 0.99, 0.63 and 0.83 arcsec), which is expected 
since their optical offsets are known as large as 2.93, 1.93, 1.20 and 1.00 arcsec, respectively.
It demonstrates that checking for the shift of 
photometric center appears effective in recognizing a candidate outburst located 
larger than 1\arcsec\ from the nucleus. 
%In fact, the SNe MIR luminosity are the lowest in our sample,
%we suspect that the constrains for other objects are even better.}

\begin{figure}
\epsscale{1.1}
\plotone{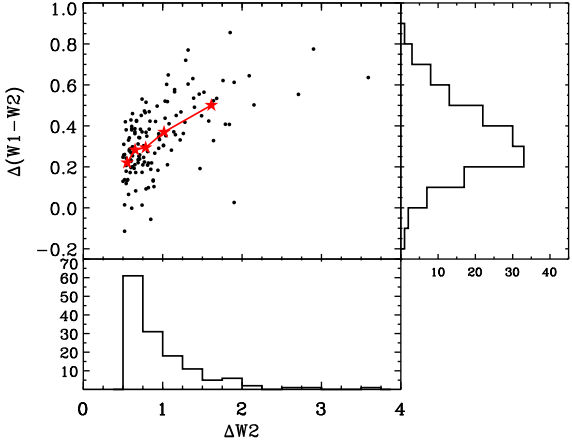}
\caption{
The color variation $\Delta\rm(W1-W2)$ versus W2 variability amplitude ($\Delta$W2) 
of the MIR outburst galaxies.
The histogram of $\Delta\rm(W1-W2)$ and $\Delta$W2 are shown in the 
right and bottom panels, respectively.
We have divided the whole sample into five equal-sized subsample sorted by $\Delta$W2
and plotted the median $\Delta\rm(W1-W2)$ of each subsample as red five-point stars.
The MIR variability shows a overall trend of redder when brighter.
}
\label{w12color}
\end{figure}

\begin{figure}
\epsscale{1.1}
\plotone{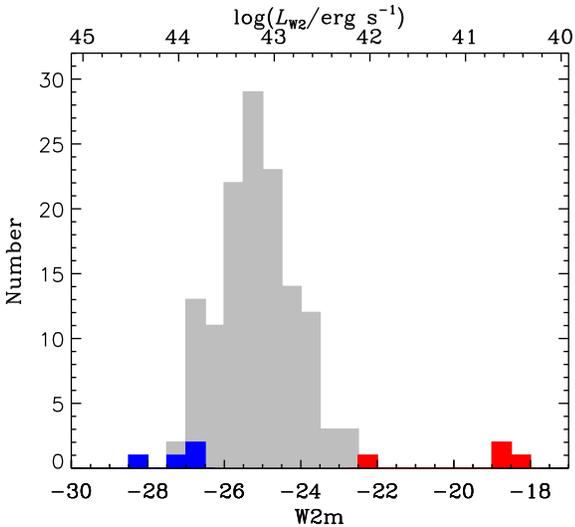}
\caption{
The histogram of peak absolute W2 magnitude.
We plot the confirmed SNe in red and radio-loud AGNs in blue.
}
\label{w2peak}
\end{figure}

\subsection{The MIR Light Curves}

Since bulk of our MIR flares have no informed optical counterparts, 
we instead try to characterize their properties by the MIR light curves themselves. 
The whole sample has brightened by average 0.63 (median 0.53) and 0.96 (0.79) 
magnitudes in W1 and W2, respectively, indicating that the 
variability amplitude at W2 band is overall larger than W1.
In other words, the W1-W2 color displays a trend of redder-when-brighter 
(RWB, see Figure~\ref{w12color}) with a median $\Delta\rm(W1-W2)=0.33$.
Only two Seyfert galaxies (SDSSJ0811+4054, SDSSJ1029+2526)
show the color changes in an opposite way.
The RWB evolution is consistent with the scenario found in CL AGNs, indicating 
a higher hot dust contribution than the starlight-dominated 
quiescent state when the AGN activity becomes stronger
(e.g., \citealt{Sheng2017,Yang2018}).

\begin{figure*}
\epsscale{0.9}
\plotone{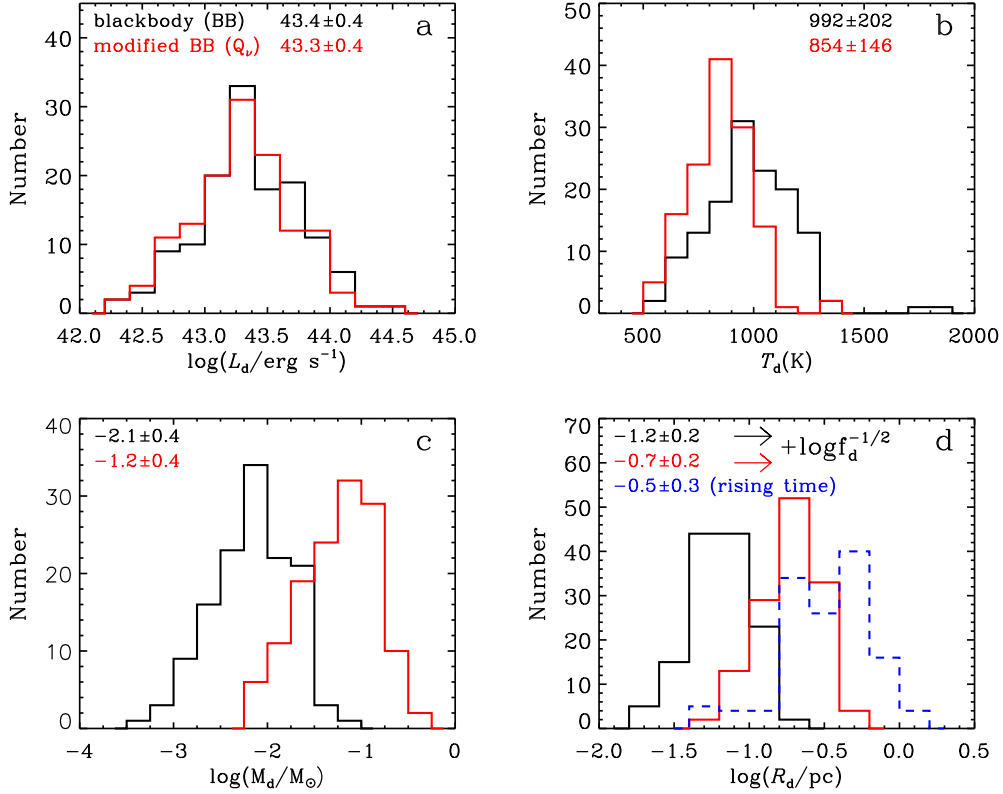}
\caption{
The histograms of the fitted dust luminosity (panel $a$), temperature ($b$), 
mass ($c$) and distance to the heating source ($d$), respectively.
All parameters are derived at the W2 maximum epoch.
The black line represents the pure blackbody case while the red shows the
fitting with dust absorption efficiency taken into consideration.
The numbers indicated in the
panels are medians and standard deviations of the distributions.
The four SNe sources have not been shown in the plots as they are 
outliers in the low end whose values can be retrieved in Table~\ref{tb_dust}.
For the dust distance (panel $d$), we have also overplotted the distance given
by the rising time scale of the \wise\ light curves (blue dashed line).
The right pointing arrows indicate that the black and red histograms 
will shift by $\rm logf_d^{-1/2}$ if the real dust
covering factor ($\rm f_d$) is not unity.
}
\label{dustpar}
\end{figure*}

In order to investigate further the properties of the MIR outbursts,
we subtracted the background emission from the light curves in the flare state (Section 2.4). 
The background-subtracted W2 absolute magnitude ($\wmax$) 
is presented in Figure~\ref{w2peak}, with the majority (95.6\%, 131 out of 137) 
ranging from $-22.5$ to $-27$, corresponding to the 
logarithmic monochromatic luminosity (\lw) between 42 to 44.
It is conspicuous that while the low-luminosity outliers ($\wmax\gtrsim-22.5$ 
or \lw~$\lesssim42$) are all SNe, the radio-loud AGNs (see \S\ref{Jet}) 
dominate at the high-luminosity end.
We stress that the minimum absolute magnitudes (or maximum luminosity)
do not always tell us the true peak since some of them are still rising in the light curves. 
If we naively suppose that the MIR outburst started from 
the middle between the first brightening data point and the quiescent state,
median rising timescale to the (current) peak is 429 days, or 398 days in the
rest-frame of galaxies.

\subsection{Dust Properties}
\label{dust}

In this subsection, we try to acquire more physical quantities by
fitting the MIR emission with dust thermal emission scenario. 
For dust grains with size distribution $N(a)$ ($a$ is the raidus of grain sphere),
density $\rho$ and absorption coefficient $Q_\nu$ at luminosity distance $d_L$,
the observed monochromatic flux at a given frequency can be calculated as below
(see also \citealt{Wang2019}).

\begin{align}
	f_\nu &=\frac{1}{4\pi d_L^2}\int_{a_{\rm min}}^{a_{\rm max}} N(a) 4\pi a^2 Q_\nu(a) \pi B_\nu(T)~da\nonumber \\
     &=\frac{3M_d}{4\rho d_L^2}\left<\frac{Q_\nu}{a}\right> B_\nu(T) \nonumber \\
\end{align} 

where 
\begin{equation}
\left<\frac{Q_\nu}{a}\right>=\frac{\int_{a_{min}}^{a_{max}}N(a) a^3(Q/a) da}{\int_{a_{min}}^{a_{max}}N(a) a^3 da};
\end{equation}

\begin{equation}
M_{\rm d}=\int_{a_{\rm min}}^{a_{\rm max}} N(a) \rho \frac{4}{3}\pi a^3 da
\label{bb}
\end{equation}

For simplicity, we assume that the dust grains follow a MRN size distribution
(\citealt{Mathis1977}; see also \citealt{Draine1984}) as
$N(a)\propto a^{-3.5}$ with $a_{\rm min}=0.01\mu$m, $a_{\rm max}=10\mu$m
and an average density of $\rho=2.7\rm g~cm^{-3}$ for silicate grains.

We begin the fit with blackbody model ($Q_\nu=1$).
However, the real dust emission is not perfect blackbody and the 
absorption coefficient should be considered.
We adopt the silicate absorption coefficients 
in \citet{Laor1993}, which gives $\left<\frac{Q_\nu}{a}\right>$=0.214
and 0.177 in W1 and W2, respectively.
%that is $Q_{\nu}\propto \nu^{1.40}$.
The fluxes in W1 and W2 are then fitted with the modified blackbody model 
to derive the dust temperature ($T_d$) and mass ($M_d$).
The luminosity is comparable to that obtained with a simple blackbody model, 
while the $T_d$ is systematically lower and $M_d$ is about 
one order of magnitude higher (see Figure~\ref{dustpar}).
We note that the $T_d$ are all below 1500~K, in agreement with 
the suppression of the sublimation temperature of
silicate and graphite grains (\citealt{Barvainis1987,Mor2012}).
All the blackbody parameters to describe the dust emission at the epoch of luminosity maximum
can be found in Table~\ref{tb_dust} and their distributions are
presented in Figure~\ref{dustpar}.

We then attempted to estimate the distance of dust emission ($R_d$)
to the central radiation source. 
By assuming spherical symmetry for the dust distribution   
%Lacking the information of 
%As the dust distribution is uncertain, assumed spherical symmetry for simplicity. 
 the distance of dust emission can be expressed by
\begin{equation}
	R_d=\left(\frac{L_{\rm d}}{4\pi\sigma T_d^4}\right)^{1/2},
\label{rdust}
\end{equation}
%This is the dust emission re-emitted from a spherical surface with radius $R_d$. 
As the dust distribution is uncertain, the above estimation may be 
oversimplified but can be treated as a strict lower limit on the scale of dust distribution (median value of 0.06~pc).  
In reality, the dust might not fully cover the central radiation source.
If the dust covering factor is $f_d$ (with unity as complete coverage),
the corresponding distance should be increased by a factor of $f_d^{-1/2}$.
Likewise, the $R_d$ for the case of modified blackbody model would be also larger, 
yielding a median value of $0.20~pc$ ($f_d=1$) that is scaled by the dust mass.
Alternatively, the MIR rising time scale can be also used as a
distance indicator of the dust responsible for the peak emission.
%since the time lags between MIR emission and central outburst are unavailable.
In this way, we obtain a median value of $0.34~pc$ 
given the median rest-frame rising time scale of 398 days.
$f_d=0.04$ (or $f_d=0.34$) from the blackbody (or modified blackbody) 
fit can generally reproduce the observed rising time scale.
%In spite of large uncertainties due to the missing details,
Above all, we may conclude that the dust is located at the order of $0.1~pc$
(see distribution in panel $d$ of Figure~\ref{dustpar})
with covering factor at the order of 0.1.

\begin{figure}
\epsscale{1.1}
\plotone{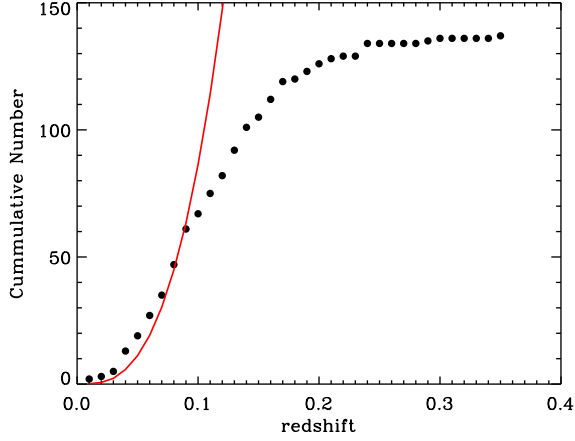}
\caption{
The accumulated redshift distribution of our sample. The curve
shows the comoving volume of SDSS footprint at the corresponding redshift
multiplied by $1.2\times10^{-6}$~$\rm Mpc^{-3}$.
}
\label{flrate}
\end{figure}

\section{Event Rate and Luminosity Function}
\label{erlf}

\subsection{Event Rate Estimation}
\label{evrate}

Regardless of their physical nature, we try to estimate the event rate of 
the MIRONG selected by us with the same manner used in \citet{Wang2018}.
The accumulative number is roughly proportional to the comoving volume 
up to $z\sim0.09$, then the increase 
rate becomes slower and finally flattens (see Figure~\ref{flrate}).
It suggests that the sample can be taken as almost complete at $z<0.09$ 
but obviously under-representative at $z>0.09$ if there is no evident
redshift evolution in the event rate.
There are 61 objects at $z<0.09$, yielding a density of 
$1.2\times10^{-6}~\rm Mpc^{-3}$ in the SDSS sky region.

%The completeness within z=0.09 can be also verified from the other
%point of view.
On the other hand, we note that the SDSS main galaxy surveys are 
originally designed to target galaxies with dereddened $r$-band magnitudes brighter than 17.77 
(\citealt{Strauss2002}).
The cut ensures that galaxies with $r$-band absolute magnitudes of $M_r<-19.5$ 
are complete, which corresponds to the luminosity range for most galaxies in our sample
(Figure~\ref{hostcm}).
Considering only for galaxies with $M_r<-19.5$ at $z<0.09$ will result in 
a density of $2.2\times10^{-4}~\rm gal^{-1}$.
Since our sample selection requires a brightening phase in the \neowise\ stage 
(spanning 5 years) and average rising timescale of about one year, 
the event rate should be divided by a factor of four to obtain the rate per year.
The final event rate is about $5.4\times10^{-5}\rm~gal^{-1}~yr^{-1}$ 
and the corresponding density rate is
$3.0\times10^{-7}~\rm Mpc^{-3}~yr^{-1}$, that is basically
consistent with $10^{-7}~\rm Mpc^{-3}~yr^{-1}$ given by \citet{Wang2018}.

\subsection{Peak Luminosity Function}

\begin{figure}
\epsscale{1.1}
\plotone{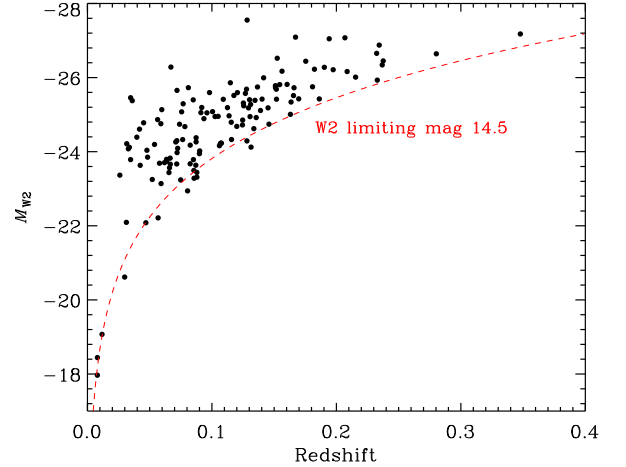}
\caption{
The host galaxy W2 absolute magnitudes (quiescent state) of our sample,
which decreases with the redshift.
The red dashed line denotes the limiting magnitude 14.5.
}
\label{zw2}
\end{figure}

We use the $V/V_{max}$ method (\citealt{Schmidt1968,Eales1993}) 
to calculate the MIR peak luminosity function of the outburst. 
The formula is adopted as below
\begin{equation}
\Phi(L)d L = \sum_{L_i \in [L-\Delta L/2, L+\Delta L/2]} \frac{1}{V_{max} f_{comp}}.
\end{equation}
Here $V_{max}$ is the maximum comoving volume within which the galaxy 
of interest is detectable at a given depth. %and selected into the sample given the depth.
Taking 17.77 and 14.5 as the limiting magnitude at $r$-band and W2, respectively, 
we computed the $V_{max}$ for each target.
In its simplest form, $\Phi(L)$ is the sum of $1/V_{max}$ 
for all objects in each luminosity bin. 
The limiting magnitude of 14.5 at W2 band is determined from our cut $W2min<14$ and a 
brightening amplitude of $>0.5$ mag (see \S\ref{mirvar}),
which is verified by the magnitude-redshift distribution of 
the final MIRONG sample (see Figure~\ref{zw2}).
$f_{comp}$ is the completeness for sources locating in the SDSS footprint 
that have been spectroscopically observed.
The spectroscopic effective area of the SDSS DR14 catalog is 9,376 $\rm deg^2$
(the full sky is 41252.96 $\rm deg^2$) and
about 92.8\% of the galaxies at $r<17.77$ have been included in the main 
galaxy sample (\citealt{Laoz2018}), thus $f_{comp}=\frac{0.928\times9376}{41252.96}=0.211$.

\begin{figure}
\epsscale{1.1}
\plotone{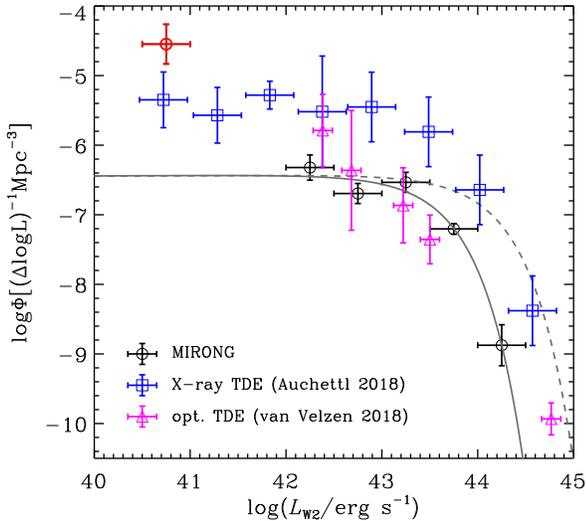}
\caption{
The W2 peak luminosity function of our MIRONG sample. 
The data points are plotted with \lw\ ranging from 40.5 to 44.5 (bin 0.5). 
The faint end (\lw$\in [40.5, 41]$) is totally contributed by
extremely low-redshift SNe (red circle). 
The high end (\lw$>42$) is fitted by a single Schechter function 
(grey line).
The shifted grey dashed line depicts the luminosity function of the primary 
emission which causes the  MIR outburst assuming a dust covering factor of 0.3.
We have also overlaid the observed 
X-ray (blue squares, data from Fig. 6 of \citealt{Auchettl2018}) 
and optical (magenta triangles, data from Fig. 1 of \citealt{vV2018}) 
luminosity functions of TDEs for comparison.
}
\label{fllf}
\end{figure}

The luminosity function in log-log space with \lw\ bin 0.5 
is shown in Figure~\ref{fllf}.
We estimate the statistical errors using the bootstrap method.
To this end, we generate $N=1000$ bootstrap samples, each of which consists of 
objects picked randomly from the original sample.
%, which allows multiple selections of individual objects. 
The errors are then given by the standard deviation in the distribution of $\Phi$
measured from the bootstrap samples. 
The Schechter function is typically used to characterize galaxy 
luminosity functions (\citealt{Schechter1976}) in the form of
\begin{equation}
\Phi(L)d L = \left (\frac{\Phi^*}{L^*}\right ) \left (\frac{L}{L^*}\right )^\alpha e^{-L/L^*} d L  ~~,
\label{eqn:schechter}
\end{equation}
where $L$ is galaxy luminosity, $L^*$ is the characteristic luminosity 
where the power-law form of the function cuts off and 
the parameter $\Phi^*$ is the normalization.  We try to fit 
the luminosity function at the high end with a single Schechter function, 
yielding log$\Phi^*=-2.92$, log$L^*=43.50$ and $\alpha=-0.05$. 
The sources at the very faint end of \lw\ (<41) 
are occupied by SNe and appear as a distinct population from the high end, 
thus we have ignored them in the fitting.
The luminosity function drops quickly at the very high end.
The flattening at \lw$\lesssim43$ could be due to the selection effect 
that requires a brightening amplitude greater than 0.5 mag.
It is possible that there are more fainter outbursts with 
relatively low brightening amplitudes (<0.5 mag), but missed by our selection.

\section{Host Galaxy and Central Black Hole}
\label{host}

\subsection{Host galaxy properties}

\begin{figure}
\epsscale{1.1}
\plotone{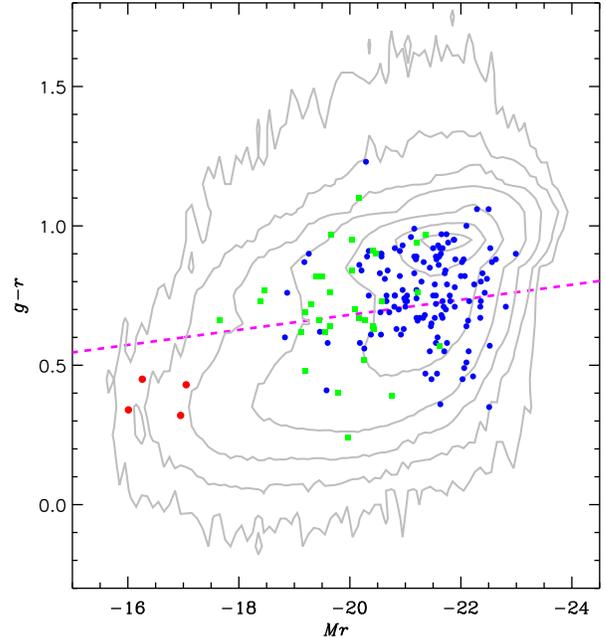}
\caption{
The color-magnitude diagram ($g-r$ vs. $M_r$) of the MIRONG sample 
(blue filled circles). 
The magnitudes and color are displayed after correction for 
Galactic dust extinction and $k$-correction to $z=0.1$.
We denote the four SNe host galaxies in red.
Contours show the density of low-redshift (z<0.2) SDSS DR7 spectroscopic galaxies.
The contour lines correspond to 5, 25, 100, 400, 2000, 5000 and 8000 galaxies 
per bin of $\Delta(g-r)=0.1$ and $\Delta M_r=0.1$.
The magenta dashed lines is the assigned line of demarcation between the blue 
clouds and red sequence. We have also plotted the optical TDEs 
(\citealt{vV2020}) for comparison (green squares), in which the magnitudes are from 
either SDSS or PanSTARRS. 
}
\label{hostcm}
\end{figure}

The host galaxy property is crucial to understand
the nature of MIRONG.
It is well known that galaxies show a bimodal distribution in the 
colour-magnitude diagram (CMD), which are mainly clustered into a red sequence and
blue cloud (e.g., \citealt{Strateva2001,Bell2004}) 
with a green valley in between.
We retrieved the apparent $ugriz$ Petrosian magnitudes from the SDSS DR14
and then corrected for the extinction using the dust maps of \citet{Schlegel1998}.
For a fair comparison of galaxies at different redshifts in the CMD, 
we applied $k$-corrections to the observed magnitudes to $z=0.1$ 
(close to the median redshift of our sample) using the 
$\tt IDL$ code $\tt KCORRECT$ (v4.3)~\footnote{http://kcorrect.org/} given by
\citet{Blanton2007} (see also \citealt{Blanton2003}).
To divide the SDSS galaxy sample into the red and blue class, 
we used the following magnitude-dependent color cut:
\begin{equation}
  g-r=-0.027*M_r+0.14
\label{cmcut}
\end{equation}
Our sample concentrates on the densest region of the CMD diagram 
(see Figure~\ref{hostcm}).
There are 82 galaxies categorized into the red sequence and the fraction (59.9\%)
is somewhat comparable to the ensemble SDSS spectroscopic galaxy sample (63.0\%).
The difference in $M_r$ is also tiny, with our sample only 0.24 magnitude brighter.
In other words, the host galaxies of the MIR outburst are not significantly
biased in terms of optical color and magnitudes. 
For comparison, the host galaxies of optically-selected TDE are much 
less luminous and most are dwarf galaxies.

\begin{figure*}
\epsscale{0.8}
\plotone{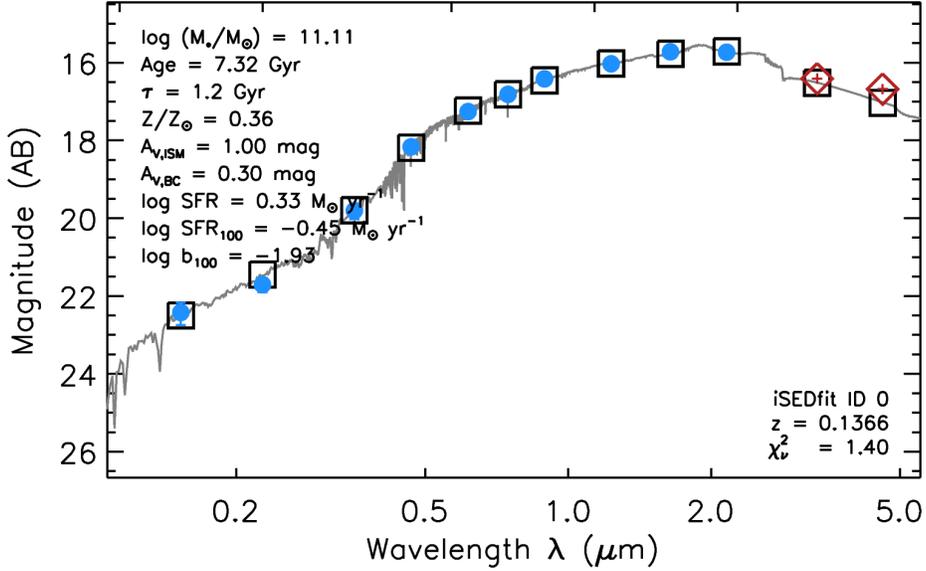}
\caption{
The SED fit of one galaxy (SDSSJ0000+1438) in our sample.
We have collected the NUV (GALEX), optical (SDSS), NIR (UKIDSS or 2MASS)
and MIR (\wise) magnitudes and then performed the fitting with {\tt iSEDfit}.
The blue filled circles are actual photometry data used in the fitting and the 
black open squares are predictions of the fitting model at the input bands.
The \wise\ W1 and W2 bands (red open diamonds) have not been used in the 
fitting but only plotted as a comparison since the PSF magnitudes
will underestimate the flux of extended sources (e.g., nearby galaxies).  
Most of the photometry errors are tiny and thus are not visible in the plots.
}
\label{sedfit}
\end{figure*}

The SDSS data release provides also value-added catalog of the galaxy intrinsic 
properties\footnote{https://www.sdss.org/dr15/spectro/galaxy/}.
For instance, the Portsmouth group has performed stellar kinematics and 
emission-line flux measurements (\citealt{Thomas2013}) 
using the publicly available codes 
Penalized PiXel Fitting (pPXF, \citealt{Cappellari2004}) and 
Gas and Absorption Line Fitting (GANDALF v1.5; \citealt{Sarzi2006}).  
GANDALF fits stellar population and Gaussian emission line 
templates to the galaxy spectrum simultaneously to separate stellar continuum and 
absorption lines from the ionized gas emission. 
Stellar kinematics are evaluated by pPXF where the line-of-sight 
velocity distribution is fitted directly in pixel space. 
The fits account for the impact of diffuse dust 
in the galaxy on the spectral shape adopting a \citet{Calzetti2001} extinction curve. 
Outputs from this fitting process include stellar velocity dispersions (\vdisp), 
emission-line fluxes, equivalent widths and BPT classifications.
Note that for one source, SDSSJ1422+1609, the galaxy parameters are not available 
in the catalog due to the lack of SDSS spectrum.  

The Portsmouth group has also provided the measurements of stellar mass (\mstar) 
through SED fitting with stellar population models (\citealt{Maraston2013}).
However, their fittings only considered the SDSS optical photometry 
and may induce bias due to the narrow range of wavelength coverage.
Therefore, we tried more comprehensive SED fittings by including 
the near IR and ultraviolet (UV) photometry, which is capable to better trace 
the old and young stellar populations, respectively.
The UV data are taken by the \emph{Galaxy Evolution Explorer} 
(\emph{GALEX}; \citealt{Martin2005}) with the near-UV (NUV) and far-UV (FUV) 
filters. 
For the NIR data, we adopted the $J, H, K$-band Petrosian magnitudes 
given by the UKIRT InfraRed Deep Sky Surveys (UKIDSS; \citealt{Lawrence2007}).
For sources that are located outside the UKIDSS footprint, 
we used the magnitudes from the Two Micron All Sky Survey 
(2MASS; \citealt{Skrutskie2006}).

After gathering the UV, optical and NIR magnitudes as well as their errors, 
we begin the broadband SED fitting utilizing {\tt iSEDfit}\footnote{http://www.sos.siena.edu/~jmoustakas/isedfit/},
which is a code to determine the \mstar, SFRs, and other physical properties 
of galaxies within a simplified Bayesian framework (\citealt{Moustakas2013}).
The fitting results are quite good (see an example in Figure~\ref{sedfit}) and the 
resulted \mstar\ are higher than those given by Portsmouth group by a median of 0.16 dex,
indicating a fraction of old stellar population probably missed by the SED fitting using only the 
optical data.
Generally, the \mstar\ distribution is consistent with the $M_r$.
Most galaxies ($92.0\%$) have stellar masses in the range $10^{10-11.5}$~\msun\ (median $10^{10.7}$~\msun,  
see Figure~\ref{mbhms}). 
The least massive five galaxies with a mass lower than $10^{9.5}$~\msun\, 
including the four SNe hosts, could be classified as dwarf galaxies 
since their masses are even lower than the Large Magellanic Cloud 
($2.7\times10^9$~\msun, van der \citealt{vander2002}).

\begin{figure*}
\epsscale{1.0}
\plotone{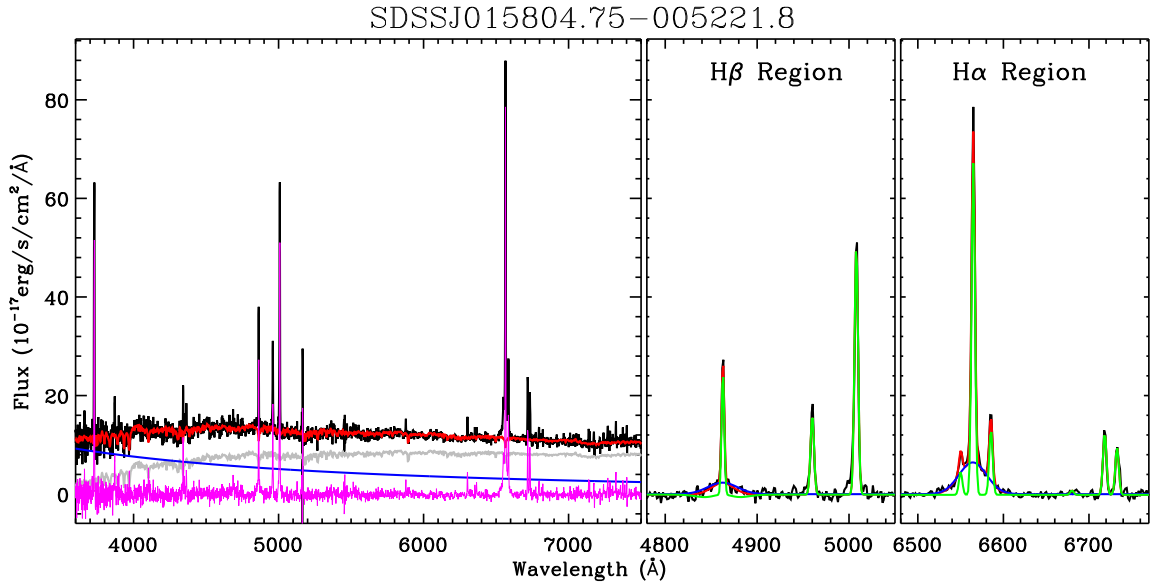}
\caption{
We show the analysis of the SDSS spectrum of SDSSJ0158-0052 as an
illustration of the spectral decomposition. The left panel displays 
the subtraction of starlight (grey) and AGN continuum (blue).
The sum of the starlight and continuum is plotted in red 
and the residual is plotted in magenta.
The middle panel highlights the Gaussian fitting of \hb-\oiii\ region, 
in which the broad \ha\ component, narrow lines and total are shown in blue, 
green and red respectively.
The right panel is similar but for \ha-\nii-\sii\ region.
}
\label{j0158}
\end{figure*}

\subsection{Nuclear Activity and Mass of SMBHs}
\label{smbh}

\begin{figure*}
\epsscale{0.8}
\plotone{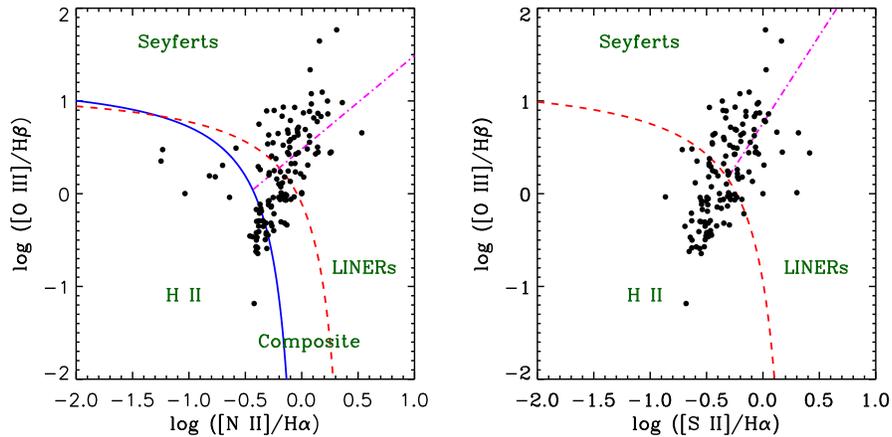}
\caption{
The locations of the MIR outburst galaxies (black filled circles) in the BPT diagram.
The lines separating the different regions are mainly drawn from 
\citet{Kewley2006} 
only with the Seyfert/LINER dividing line (magenta dashed line) in the left panel 
from \citet{Cid2010}.
}
\label{bpt}
\end{figure*}

As we have mentioned in \S\ref{intro}, 
the transient accretion onto SMBHs is a major population
of extragalactic transients associated with galaxies. 
It is thus useful to assess the pre-outburst nuclear activity of these galaxies,
which may shed light on the nature of the outburst. 
The narrow-line ratios of galaxies, namely their location 
at the BPT diagram, can be used as a diagnosis of the nuclear activity, 
but only the broad emission lines can give unambiguous evidence of active SMBHs 
and plausible estimated \mbh.
In order to detect the potential broad lines, we have performed careful spectral
analysis on the SDSS spectrum of these galaxies.

Our spectral fitting procedures are as follows: 
(1) subtracting the starlight and AGN continuum to obtain 
the emission-line residuals; 
(2) Gaussian fitting to emission lines including 
broad components when necessary.
The spectrum is corrected for the Galactic extinction with the extinction 
map of \citet{Schlegel1998} and the reddening curve of \citet{Fitz1999}. 
We model the starlight component with the stellar templates of \citet{Lu2006}, 
which were built from the simple stellar population spectra 
(\citealt{Bruzual2003}). 
The AGN continuum is modeled as a power law. 
After subtracting the starlight and AGN continuum component, 
we try to fit emission lines with multiple Gaussians while only broad components 
($\sigma>500~\kms$) for \ha, \hb, \mgii\ and \hei\ are allowed to vary. 
Although emission lines in a fraction of galaxies can be fitted with a 
broad \ha\ component superimposed on the narrow component, 
only those broad \ha\ lines with S/N higher than 10 are considered as valid, 
resulting in a final sample of 26 galaxies.
The robustness of our fitting results is demonstrated 
in SDSSJ0158-0052 (see Figure~\ref{j0158}), which is a well-known low-mass AGN 
candidate selected by broad \ha\ emission (\citealt{GH2007}; \citealt{Xiao2011}).
The \mbh\ is subsequently calculated by empirical virial mass estimator 
($\mbh=fRv^2/G$) for single-epoch spectra using the formalism 
presented in \citet{GH2007}.
This method postulates that the broad-line region (BLR) gas is virialized 
with velocity dispersion characterized by the widths of broad lines and a distance 
to the BH estimated from the conventional radius-luminosity relation 
(e.g., \citealt{Kaspi2005,Bentz2013}). 

\begin{figure}
\epsscale{1.15}
\plotone{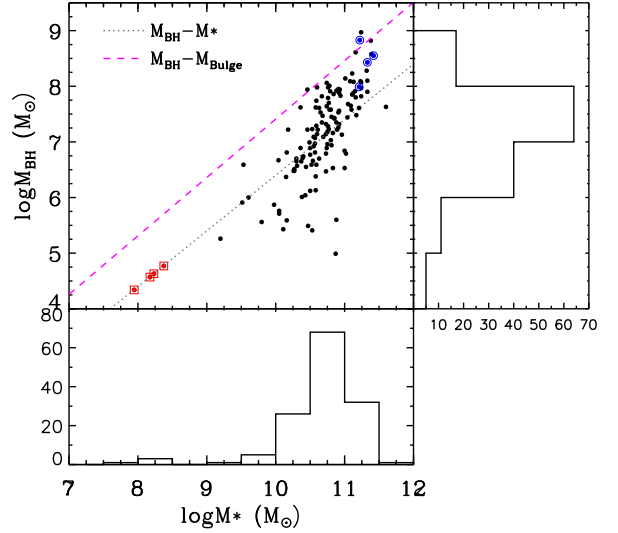}
\caption{
The black hole mass (\mbh) versus stellar mass ($M_*$) distribution of our sample. 
The red solid circles are SNe and the blue solid circles are strong jetted sources.
The histogram of \mbh\ and $M_*$ are shown in the right and bottom panels.
Overlaid in grey dotted and magenta dashed lines are the correlation between 
\mbh\ and bulge mass (\citealt{McConnell2013}) and total stellar mass 
(\citealt{Reines2015}), respectively.
}
\label{mbhms}
\end{figure}

Regarding the narrow-line sources, their nuclear activity can be alternatively 
identified by their positions in the BPT diagram (see Figure~\ref{bpt}). 
According to the classification of Portmouth group (with the broad-line
AGNs updated from our own fittings), our sample can be categorized into 
37 (14 with broad lines) Seyferts, 23(4) LINERs, 35(2) starforming galaxies 
and 41(6) composites. 
Broad-line starforming galaxies are not common but indeed exist
(see Figure 8 of \citealt{Liu2019}). 
However, the fraction in our sample looks somewhat high (2/35), which may 
indicate some of the MIR outburst in starforming galaxies are driven by AGNs.
The total number of AGNs contained in our sample is 49 (37+4+2+6)
when taking count of both Seyferts and broad-line sources in other BPT types.
No clear evidence of intense AGN activity is found for other sources. 
We caution that LINERs can be treated as weak AGNs powered by SMBHs.
In the absence of broad-lines, we have to estimate their \mbh\ with 
other approaches, such as the correlations with either the \vdisp\ or mass of 
galactic bulge (\mb) established in local massive galaxies 
(see \citealt{KH2013} for a review).
Our experiences and other works suggest that the velocity dispersion
after correction of the instrument broadening ($70$~\kms) is reliable 
down to $\sim$50$-$60~\kms\ (e.g., \citealt{Zahid2016}; \citealt{Chilingarian2017}).
Hence we adopted only the \mbh-\vdisp\ relation when $\vdisp>50$~\kms,
leaving seven galaxies without \mbh\ measurements because of low \vdisp.
Although the \mbh-\mb\ relation (e.g., \citealt{McConnell2013}) has been 
extensively used to estimate \mbh, the SDSS resolution is generally too low 
to isolate the bulge component from disk. 
We thus used the relation between \mbh\ and 
total stellar mass (\citealt{Reines2015}) for the last seven objects.
Our final sample has a broad range of \mbh\ with logarithmic mass 
from 4.5 to 9.0 (median 7.3, see Figure~\ref{mbhms}).

%\shu{I suggest to simply delete this paragraph. Even at the high-mass end, 
%the estimation is crude either.}
%{It is interesting to note that for 17 galaxies the estimated 
%BH mass is lower than $10^6$\msun~, so-called intermediate-mass black hole (IMBH). 
%Nevertheless, we caution that our estimate of \mbh\ is very crude, 
%particularly, the proof of those methods in dwarf galaxies is far away from solid.
%One may not take the \mbh\ at the low-mass end (\ms$<10$) seriously 
%considering these uncertainties, which even does not necessarily mean a 
%real existence of central BH. }

\section{Nature of the MIR flares}
\label{nature}

The above analysis of the properties of MIR flares and their host galaxies,
allow us to explore further their physical nature.
MIR flares can be generally attributed to non-thermal emission from jets or dust thermal emission 
heated by different processes, such as supernova, AGN or TDE.  
We will discuss each of these possibilities below.

\subsection{Infrared Luminous Supernova}
\label{SNe}

As mentioned in \S\ref{opt}, SDSSJ0936+0615, SDSSJ1531+3724, SDSSJ1540+0054
and SDSSJ1554+1636 are known hosts of SNe reported by ASAS-SN\footnote{
http://www.astronomy.ohio-state.edu/asassn/sn\_list.html}.
Interestingly, their peak MIR luminosity appears to be lowest, 
with absolute W2 magnitudes fainter than -22 (or \lw<42).
The reason for them to pass our selection threshold of variability amplitude
($>$0.5 mag) could be the dwarfness of their host galaxies (see Figure~\ref{hostcm}), 
which makes the MIR outburst luminosity over luminous with respect to their host. 
Indeed, the four SNe galaxies are the least massive ones
and occupy the low-mass end (\ms$<9$).
In addition, these galaxies appear very young in terms of their blue colors (g-r<0.5).
Hence, the SNe in our sample, in a word, occupy the lowest MIR luminosity end, and  
reside in dwarf starforming galaxies. 

%\shu{the sentence below is vague}
%\jn{On the other hand, the SNe dominance at the extreme low-mass end 
%implies that the occurrence of SMBHs could be intrinsically scarce, meeting with 
%the fact of very few central BHs founded in galaxies at this mass range 
%(e.g., \citealt{Reines2015}).}  

%\shu{SNe is singular form or not?}

The spectroscopic follow-ups have classified the four SNe into three type~II 
and one Ia-91T like sources.
Previous studies suggest that the IR emission of SNe IIn are statistically 
more luminous and long-lasting than other types (e.g., type Ia)
due to the heating of preexisting dust in the circumstellar medium (CSM) 
%through radiative shocks between the expanding shell and the dense wind of 
%the progenitor 
(\citealt{Fox2013,Tinyanont2016}). 
The SNe Ia usually show very weak MIR emission and are not detectable in three years 
after the explosion (\citealt{Tinyanont2016}) except the Ia-CSM subclass.
In the same way, some 91T-like SNe also display the interacting CSM 
(e.g., \citealt{Harris2018}), which could be responsible for the observed MIR flare.
The luminous infrared transients uncoverevd by SPIRITS project 
are mainly obscured core-collapse SNe with peak $4.5\mu$m 
(\spitzer, roughly \wise\ W2) magnitudes between -14 and -18.2 
(\citealt{Jencson2019}), which are significantly fainter than that in our sample.

\begin{figure}
\epsscale{1.15}
\plotone{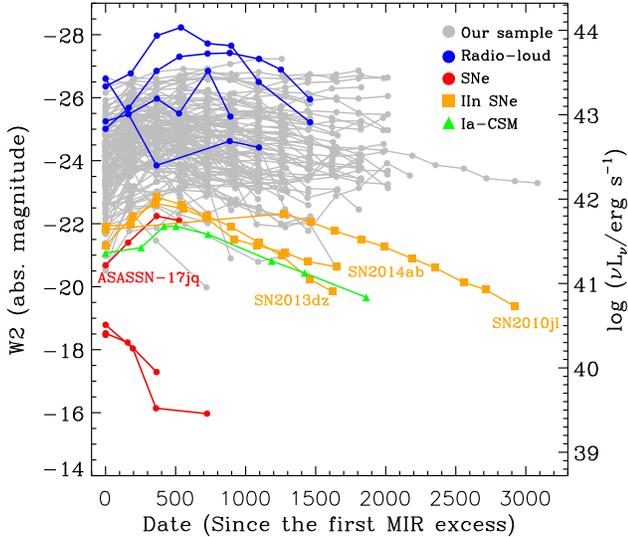}
\caption{
The time-resolved W2 magnitudes with date zeropoint set as the time of 
first MIR excess.
Our sample has been plotted in solid grey circles with the 
radio-loud sources and SNe highlighted in blue and red, respectively.
We have also overplotted other IR luminous SNe (orange squares), 
which are among the brightest known to us, for comparison. 
The Ia-CSM light curve (green triangles) are drawn from \citet{Fox2013}.
}
\label{abslc}
\end{figure}

The evolution of absolute W2 magnitudes as a function of time is presented in 
Figure~\ref{abslc}.
It is clear that the four SNe have the faintest MIR emission (see also the 
peak magnitude distribution in Figure~\ref{w2peak})
and their duration is relatively short.  
While the SNe occurred in SDSSJ1531+3724 (ASASSN-17jq) is relatively bright, 
it is still fainter than non-SNe objects.
In our previous work (\citealt{Jiang2019}), we have checked for the \wise\ light curves
of all SNe in the public catalog reported between 2008 and 2018.
Among them, the most luminous ones are all IIn (e.g., SN2010jl,
SN2013dz and SN2014ab) with luminosity $L\rm_{W2}\sim10^{42}$~\lum,
which is comparable to that of ASASSN-17jq.
In summary, the MIR luminosity of our sample is systematically 
higher than SNe by at least 2-3 magnitudes, thus the SNe scenario is
disfavored as the origin for the bulk of MIRONG.

\begin{figure}
\epsscale{1.1}
\plotone{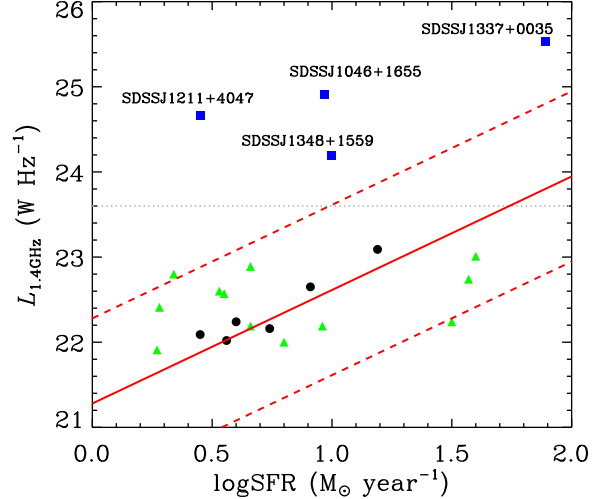}
\caption{
The 1.4~GHz luminosity versus the SFR of radio detected sources in our sample.
The red line is the expected radio emission from star formation (Equation 3 in 
\citealt{Davies2017}). One index offset from the expectation are plotted as
red dashed lines. 
We have highlighted the radio-loud AGNs with blue squares and other
AGNs (Seyferts or broad \ha\ objects) with green triangles.
The grey dotted line represents the radio luminosity beyond which
can be considered as undisputed jet origin (\citealt{Kellermann2016}).
}
\label{sfr1.4}
\end{figure}

\subsection{Non-thermal emission from Jet}
\label{Jet}

We next examine the possibility of non-thermal emission produced by relativistic jets.
It has been proved that the synchrotron radiation of jets can contribute 
significantly to the MIR emission as well as its variability 
(e.g. \citealt{Jiang2012,Liao2019}).
%If that is the case, one may wonder whether the jet in act
%is a pre-existing one or newly launched by SMBH.
%Aiming to check the radio intensity before the flares, 
We cross-matched our sample 
with the catalog of Faint Images of the Radio Sky at Twenty cm (FIRST)\footnote{http://sundog.stsci.edu/cgi-bin/searchfirst} using a matching radius of 5\arcsec, 
resulting in 22 objects detected with S/N higher than 5. 
There are eight objects outside of the FIRST footprint, so 
we matched them with the NRAO VLA Sky Survey (NVSS) catalog\footnote{https://www.cv.nrao.edu/nvss/NVSSlist.shtml} but found non-detections.

The radio detections do not necessarily suggest the association with jet activity, since
the star formation may also contribute to the radio emission.
We used the star formation rate (SFR) given by \citet{Chang2015},
%\footnote{http://irfu.cea.fr/Pisp/yu-yen.chang/sw.html}, 
which is derived
by fitting the broadband SDSS+\wise\ SED using MAGPHYS (\citealt{daCunha2008}).
%They have simultaneously and consistently modeled both the attenuated 
%stellar SED and the dust emission at 12 and 22 $\mu$m, producing robust new
%calibrations for monochromatic MIR SFR proxies.
The SFR measurements for 128 out of 137 galaxies are found.
For the remaining sources, the SFRs were estimated from the 24~$\mu$m flux
\footnote{Here we adopted the \wise\ W4 band (22~$\mu$m) flux as an approximation.} 
(\citealt{Chary2001}).
% http://david.elbaz3.free.fr/astro_codes/chary_elbaz.html
Then we calculated the expected radio flux from star formation
using the correlation between 1.4~GHz luminosity ($L_{1.4G}$) and SFR 
(Equation 3 in \citealt{Davies2017}). 
As displayed in Table~\ref{radio} and Figure~\ref{sfr1.4}, 
the radio emission for star-forming and composite galaxies 
are fully consistent with the expectation from the star formation.
Only SDSSJ1337+0035, which is a composite but shows extremely high radio power, is an exception.
We then used the radio-loudness parameter $R$ to quantify the radio intensity
for AGN sources (Seyferts or LINERs) plus SDSSJ1337+0035, which is defined as the ratio of the flux densities
between 6~cm and 4400~\AA\ (\citealt{Kellermann1989}).  
Here the 6~cm flux is derived from the 1.4G~Hz flux assuming a spectral slope $-0.7$.
The 4400~\AA\ flux is converted from the 5100~\AA\ flux, which is derived 
from the bolometric luminosity (\lbol). 
The latter is computed from the \oiii\ luminosity (\citealt{Lamastra2009}) assuming a bolometric correction of 8.1 
(\citealt{Runnoe2012}). 
The derived radio-loudness parameters are listed in Table~\ref{radio}.

Four galaxies (SDSSJ1046+1655, SDSSJ1211+4047, SDSSJ1337+0035, SDSSJ1348+1559) 
stand out from the rest in above evaluations. 
They show radio power 38-600 times higher than that predicted from SFR. 
%while the other 18 sources display a more comparable level.
Meanwhile, their radio-loudness (>1000) is at least two orders of magnitude higher
than other sources.
Moreover, the radio luminosity for the four sources is higher than 
$10^{23.6}~\rm W~Hz^{-1}$, above which a radio-loud AGN can be classified (\citealt{Kellermann2016}). 
Therefore, they are likely radio-loud AGNs, for which the MIR flares 
could originate from the non-thermal emission of pre-existing jets. 
%The remaining AGN sources, particularly those with $R>10$, 
%can still possess weak jets (denoted as 'w.Jet' in Table~\ref{radio}), 
%but the jet contribution to the MIR emission is not sig.
For those galaxies that are not radio-loud before the MIR flare, 
we cannot rule out the probability of a new-formed relativistic jet 
contributing the MIR emission. 
Unfortunately, no quasi-simultaneous radio observations are available. 
In fact, by examining the \wise\ light curves in each epoch, 
we did not found any evidence for intraday variability, which 
could indicate the presence of a relativistic jet (\citealt{Jiang2012}). 
Timely radio follow-ups would be helpful to further test this intriguing scenario.

\subsection{Dust Echo of Transient SMBH Accretion}

%39(15 with broad lines) Seyferts, 21(2) LINERs, 35(2) starforming and 41(6) composites.
After removing the 4 known SNe and 4 radio-loud sources, 
we now investigate the origin of MIR outbursts for the other 129 galaxies in our sample. 
As we analyzed in \S\ref{smbh}, there are 49 galaxies 
that show evidence of AGN activity in their optical spectra. 
%hinted by either
%the presence of broad-lines or narrow-line ratios of SDSS spectra. 
In addition, 16 galaxies can be classified as LINERs. 
Therefore, the 128 galaxies (except SDSSJ1422+0609 for which the SDSS spectrum is not available) 
can be grouped by two populations, including 
65 AGNs (49 Seyferts and 16 LINERs) and 63 quiescent galaxies.

\subsubsection{Echoes of Turn-on AGNs}

The UV/optical variability on various time scales are well known to be an 
inherent property of AGNs albeit the physical mechanism behind is not 
fully understood (\citealt{Ulrich1997}). 
The amplitude of optical variability is typically a few tenths 
of magnitudes within time-scale of months, but also showing larger
variations over longer time-scales. 
There is an increasing population of dramatically variable AGNs,
so-called CL AGNs, which exhibit flux rising/declining in continuum 
($\Delta m>1$ mag) over several years and change accordingly spectral types
(e.g., \citealt{LaMassa2015,Runnoe2016,MacLeod2016,Yang2018}).
Specifically, the CL AGNs with rising/declining photometric light curves are termed "turn-on"/"turn-off".
In addition to state changes found in CL AGNs, major outbursts
have also been found in AGNs (e.g., \citealt{Graham2017,Trakhtenbrot2019}).

In the AGN unification, the dusty torus is a vital ingredient to address 
a variety of phenomenons in different types of AGNs. 
The torus will unavoidably absorbs part of the UV-optical photons 
from the accretion disk and reprocesses them into the IR. 
% with the accurate fraction determined by the torus covering factor.
The picture has been widely accepted through the 
spectral energy distribution (SED) fitting (e.g., \citealt{Fritz2006,Nenkova2008,Stalevski2012}) 
and IR reverberation mapping (e.g., \citealt{Suganuma2006,Koshida2014,Lyu2019}).
It is plausible that our discovered MIR flares are  due to dust echoes
of turn-on AGNs. 
Such a scenario can be tested by obtaining new optical spectra 
to see whether there is evident spectral evolution.
In fact, we have performed a detailed study of SDSSJ1115+0544, which has undergone 
a brightening by 2.5~mag in $V$-band over $\sim120$ days, 
then faded by 0.5 mag over 200 days, followed by a plateau lasting for $>600$ days.
The multi-epoch optical spectra over 400 days in the plateau phase 
revealed newly formed and steady broad \ha, \hb\ emission, 
that is compatible with the characteristic of a turn-on AGN (\citealt{Yan2019}).
Intriguingly, together with the three objects (SDSSJ1554+3629, 
SDSSJ0945+4814 and SDSSJ1133+6701) mentioned in \S\ref{opt}, the four 
reported sources are all classified as LINERs in the BPT diagram, which may 
suggest a uniform class of turn-on system (\citealt{Frederick2019}).

\begin{figure}
\epsscale{1.1}
\plotone{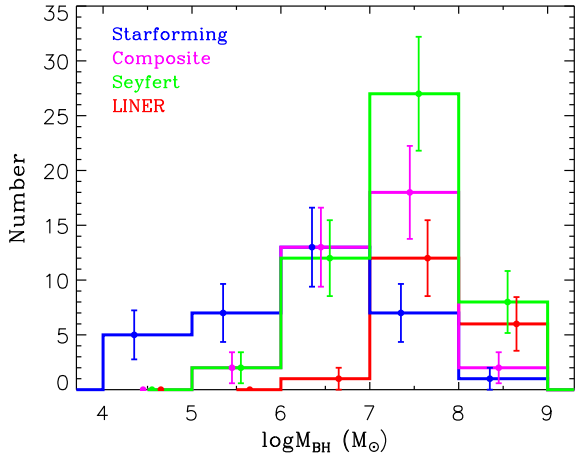}
\caption{
The histograms of \mbh\ for starforming (blue), composite (magenta), 
Seyfert (green) and LINER (red) galaxies in our sample as classified 
by the BPT diagram.  
The bars errors are the Poisson fluctuations of
the number in each bin.
Noting that we have put all broad-line sources into the 'Seyfert' subset
whatever their BPT types are.
}
\label{mbh}
\end{figure}

\subsubsection{Echoes of Tidal Disruption Events}

Nearly half objects in our sample show no signs of AGN activity, 
neither Seyfert nor LINER, in the SDSS spectra. 
The giant nuclear flares of inactive galaxies are usually attributed to TDEs.
For Schwarzschild BHs, the TDEs are observable only when the BH masses 
are lower than the Hills mass ($\sim10^8$~\msun).
Otherwise, the star will be swallowed wholly rather than produce an
electromagnetically luminous flare
because its tidal radius is within the horizon (\citealt{Hills1975,Rees1988}).
The \mbh\ distribution of our sample peaks at $2.0\times10^7$~\mbh\ with 
only 17 (12.4\%) greater than $10^8$~\msun\ (see Figure~\ref{mbhms}). 
If excluding Seyferts and LINERs, only three galaxies are more massive 
than the Hills mass (see Figure~\ref{mbh}), hence the TDE interpretation for
the MIR bursts is possible. 
%entirely %feasible as far as the BH size is concerned.

Similar to AGNs, the UV/optical photons released by TDEs can be absorbed 
by dust in the vicinity of BH and reprocessed into the IR band.
giving rise to an IR flare like an echo.
\citet{Lu2016} have calculated the light curve of the IR echo with
1-D radiative transfer model and showed that the dust emission peaks 
at MIR (3-10$\mu$m). They have predicted that the 
typical luminosity is between $10^{42}$ and $10^{43}$~\lum,
depending on the dust covering factor (ranging from 0.1 to 1), which 
is fairly comparable with  that inferred in our sample.
Such an IR echo has been detected in a handful of optical TDEs from the \wise\ data,  
(\citealt{Jiang2016,vV2016}) 
though the amplitude of IR variability is smaller ($\sim0.2$~mag) compared with 
the sample in this work.
Furthermore, \citet{Dou2016} have reported the long-lasting IR echoes from 
four coronal-line TDE candidates (\citealt{Wang2012,Yang2013}). 
These findings suggest that IR echoes might be ubiquitous for TDEs occurred in 
gas (dust)-rich nuclear environment.

On the other hand, the IR echo itself is suggestive of TDEs 
in dust-rich environment.
\citet{Wang2018} have conducted a systematical search 
and obtained a sample of 19 low-redshift (z<0.22) 
non-Seyfert galaxies that display slow decline in the MIR emission,
reminiscent of those coronal-line TDEs (\citealt{Dou2016}).
Unfortunately, because they are already in the late fading stage, 
the spectroscopic observations might be too late to identify the 
TDE-like characteristics.
In contrast, our objects are observed at much earlier phase, 
either still rising or just decaying, which are valuable  
for timely multi-wavelength follow-ups.
By checking for optical counterparts (see \S\ref{opt}), we found that 
SDSSJ0158-0052 (PS16dtm) and SDSSJ1620+2407 (ATLAS17jrp) have been indeed 
alerted as TDE candidates \citep{Blanchard2017,Fraser2017}.
The identifications of TDEs in our sample are thus encouraged. 
We caution that even for active galaxies in our sample, the TDE scenario is  
possible and worthwhile to explore.
One universal manifestation accompanied with AGN TDEs 
(e.g., \citealt{Tadhunter2017,Blanchard2017,Kankare2017}) 
is long-lasting and luminous MIR flares as a result of echoes of dusty torus 
\citep{Dou2017,Jiang2017,Jiang2019}.

Given the faintness in the optical bands for most MIRONGs (\S\ref{opt}), 
they could be promising candidates of dust-obscured TDEs 
that are still poorly explored. 
If not due to obscuration, their optical emission should be otherwise 
intrinsically weak, which is especially possible in the TDE scenario.
The electromagnetic output of TDEs was initially thought to be dominated        
by X-ray or FUV emission from accretion, making the origin of 
optical emission remains hotly disputed (\citealt{Roth2020}).
Therefore, some objects in our sample could belong to classical TDEs, which are 
luminous in X-ray/FUV but are faint in optical band.
In this sense, the IR search of TDEs is superior to other individual bands 
as long as the BH is set in dusty environment.

\begin{figure*}
\epsscale{1.0}
\plotone{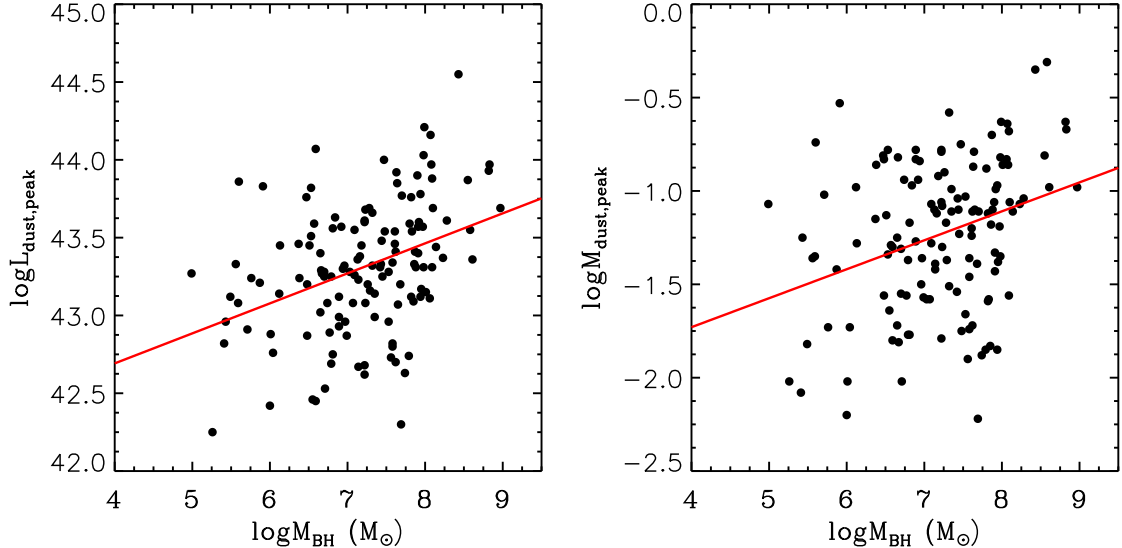}
\caption{
Left panel: the correlation between peak MIR luminosity ($L_d$) and \mbh.
Right panel: the correlation between the inferred dust mass ($M_d$) and \mbh.
Overlaid in red are the linear least-squares fitting.
The $L_d$ and $M_d$ values are derived by dust with absorption efficiency
(see Table~\ref{tb_dust}).
}
\label{lirmbh}
\end{figure*}

Based on the sample of 19 galaxies, \citet{Wang2018} have found that the peak MIR luminosity is well 
correlated with the \mbh. However, the correlation (excluding SNe) appears less significant in our sample presented 
here (see Figure~\ref{lirmbh}), though a Spearman rank analysis gives a correlation coefficient
of $\rho= 0.377$, corresponding to a null hypothesis probability of $p=8\times10^{-6}$.
If we examine the dust mass instead of luminosity, 
the correlation is even less obvious ($\rho= 0.116$, $p=0.001$).

\section{Discussion}

%We have looked into the physical nature of the MIR outburst above. 
Similar to \citet{Wang2018}, our tentative conclusion is that the 
MIRONGs discovered by us are most likely associated with transient SMBH accretion, 
either in the form of TDE or turn-on AGN activity. 
Only a small fraction of them are due to IR-luminous SNe or non-thermal 
emission from relativistic jets. 
The conclusions are supported by several lines of evidence:
1) the outburst locations prefer to be in the region very close to
the galactic center; 2) the maximum MIR
luminosity is higher than the brightest SNe by 1-2 orders; 3) 
only few have radio-loud jet emission before the outburst;
4) the luminosity function broadly agrees with results of optical
and X-ray TDEs (or nuclear transients). 

In order to differentiate the diverse nature of these MIRONG,
multi-wavelength follow-up observations have been conducted,
including optical spectra (Y. B. Wang et al. in preparation),
NIR photometry (H. Liu et al. in preparation), radio (\citealt{Dai2020})
and X-ray observations.
So far, we have successfully obtained a sizable sample of likely TDEs
or turn-on AGNs, which will be reported in the following series of papers.
These events are generally not detected in the optical band and
thus largely overlooked by previous works, hence present a significant advance 
over previous studies of similar events.
%Even for the SNe and jet cases, the unique sample may give us
%useful insights into the MIR emission of SNe and the jet flicking.

\subsection{TDE Demography and Missing Energy Puzzle}
\label{obsTDE}

The observations of TDEs are initially restricted to the serendipitous discoveries 
of X-ray flares in the centers of galaxies, yielding a very limited number of 
candidates.  The number has grown remarkably in the past decade 
with the advent of dedicated optical time-domain surveys 
(\citealt{Komossa2015,vV2020}). 
However, there is still a large discrepancy in the event rate between 
observation and theory (\citealt{Stone2020}). 
The observations usually find a rate of $\sim10^{-5}~{\rm gal}^{-1}~{\rm yr}^{-1}$
(\citealt{Donley2002,vV2014}), 
which is at least an order of magnitude 
below the theoretical estimation in realistic galactic nuclei 
(\citealt{Wang2004,Stone2016}).
Although dynamical mechanisms to suppress TDE rates have been proposed, 
they seem unlikely to work in practice (e.g. \citealt{Lezhnin2015}). 
One solution may be a very broad intrinsic TDE luminosity function of 
which we have so far only seen the high end at given band 
(e.g., \citealt{Blagorodnova2017,vV2018}) in view of the 
complicated radiative emission mechanisms (\citealt{Roth2020}).
The TDE demography at any single band can be thus seriously underestimated. 

Moreover, the rate discrepancy becomes more specific and definitive when 
optical TDEs are found being overrepresented in post-starburst 
(or so-called E+A) galaxies with the rate 
elevated by a factor of $\sim100$ (e.g. \citealt{Arcavi2014,French2016}).
This preference can not be fully accounted for by selection effects
in comparison with control samples (\citealt{Law-Smith2017}).
Such an enhancement is difficult to understand in theory, even though
their hosts show high central stellar densities (\citealt{French2020}).  
This would require violent dynamical processes to occur but it 
seems impossible in most post-starburst galaxies.

Another observational discrepancy is known as the missing energy puzzle, namely 
a huge shortage of the observed energy relative to predictions.
In the simplest picture, if half stellar material is accreted onto the BH, 
the expected energy released is $\sim$$10^{52}${--}$10^{53}\,$erg. 
The observed total energy of optical TDEs is only $\sim 10^{51}$~erg.
Where is the missing 90{--}99\% of the energy?
Some models argue that the real accretion efficiency is actually quite low
due to advection or outflow (e.g., \citealt{Svirski2017,Dai2018}).
\citet{Lu2018} suggest that the missing part may be in the 
unobserved extreme-UV (EUV) band and/or in the form of relativistic jets. 
The jet mode is disfavored as the rareness of jetted TDEs
but the EUV option coincides with the theoretical 
peak and the Rayleigh-Jeans-like optical/NUV SED.
If that is true, at least in several TDEs, the issue then goes to 
how to measure the EUV energy which is invisible directly.

Two recent studies may shed light on an avenue to solve or 
alleviate the puzzles above.
\citet{Tadhunter2017} has reported a candidate TDE in F01004-2237 
from a sample of 15 nearby ultra-luminous infrared galaxies (ULIRGs), 
suggesting an event rate as high as $\sim10^{-2}~{\rm gal}^{-1}~{\rm yr}^{-1}$.
The ULIRGs are generally starburst systems with a large amount of dust
which is able to even veil energetic outbursts like TDEs.
In the meantime, the energy absorbed by dust will manifest the 
orginal TDE emission as a luminous IR echo.
The picture is nicely corroborated by the huge MIR flare detected in 
F01004-2237 (\citealt{Dou2017}), whose integrated IR energy till now is 
$>3\times10^{52}$~erg.
The other work claims a dust-enshrouded TDE in the western nucleus (B) 
of the merging galaxy pair Arp~299 (\citealt{Mattila2018}).  
This event is almost silent in optical and X-ray, but displays long-lasting 
echoes in the NIR and MIR with a total radiated energy $>1.5\times10^{52}$~erg.
The missing energy puzzle is almost settled in both systems when the IR energy
is taken into account.

The two case studies also hint that perhaps the TDEs in post-starburst systems 
are only the tip of the iceberg and the real TDE rate in 
starburst galaxies could be immense (\citealt{Guillochon2017NA}). 
It is widely believed that the galaxy mergers or interactions are
capable of driving galactic-scale gas inflow to the nuclear region
and triggering intense star formation.
Even for an isolated disk galaxy, its center could be a concentrated region of 
gas and dust ornamented by enhanced star formation.
Indeed, a considerable fraction of galaxies in the MIRONG sample are
starforming systems as indicated by their SDSS fiber spectrum, whose nucleus 
are easily obscured as a result of intense star formation process.

Interestingly, the MIRONG event rate estimated by us
is $5.4\times10^{-5}~{\rm gal}^{-1}~{\rm yr}^{-1}$ (see \S\ref{evrate}), 
 which is comparable to that of optical TDEs.  
It would be of a high value if a large percentage of them
are eventually confirmed as real TDEs. 
Furthermore, if we naively assume a typical dust covering factor 
as AGN torus, the inferred luminosity function of the MIRONG 
is comparable with the observations
of X-ray (\citealt{Auchettl2018}) and optical TDEs (\citealt{vV2018})
at high end (see Figure~\ref{fllf}).  
The faint-end discrepancy is at least partly caused by the imperfect correction 
for the sample completeness of various selection approaches.   
We emphasize again that our selection criteria is somewhat strict
(>0.5 mag brightening). In the future we will extend the sample by a relaxed 
requirement to obtain a more complete view of the event rate
and the luminosity function at faint end. 
The new population that is largely missed by optical surveys, due to
either dust obscuration or intrinsically faintness, may offer us a 
promising approach to solve the problems of TDE demographics and missing energy.

\subsection{Implication on the Turn-on and Duty Cycle of AGNs}

CL AGNs have gradually become a rather popular phenomenon. 
The early searching has been focused mainly on known quasars, 
which naturally yield out more "turn-off" AGNs. 
The systematical discovery of "turn-on" cases have to start from a  
much larger galaxy sample, consisting of both Seyfert and normal galaxies. 
Galaxies which show rapid transformation from a quiescent galaxy to a type~1 AGN 
within several months to years are of great interest. 
So far, only two unambiguous systems (iPTF~16bco and SDSSJ1115+0544) 
have been reported (\citealt{Gezari2017,Yan2019}).
We refer them to \emph{bona fide} turn-on AGNs, which are 
different from the normal CL AGNs that usually changed from 
type~2/1.9/1.8 to type~1. 
Some CL AGNs have shown to change their types 
back and forth frequently (e.g., \citealt{Denney2014,McElroy2016,Oknyansky2019}), 
implying the existence of a persistent but 
unstable accretion disk, e.g., susceptible and sensitive to gas feeding. 
In comparison, the \emph{bona fide} turn-on AGNs can serve as a more genuine
laboratory than normal CL AGNs to explore the ignition mechanism of SMBHs, e.g.,
the rapid formation of accretion disk as well as the origin of 
concomitant multiwavelength emission (e.g., X-ray, radio).

It is worthwhile to note that both iPTF~16bco and SDSSJ1115+0544 are 
present in our MIRONG sample.
It suggests that more similar systems would present in our sample 
in view of the quiescence of many galaxies before the outburst, and 
the selection of "turn-on" AGNs might be effectively using the MIR 
light curves. 
% looks surprising 
%as it is absolutely subject to the reprocessed emission of dust. 
The dust covering factors ($f_d$) of quiescent SMBHs are around $10^{-2}$ 
as revealed by the IR echoes of optical TDEs, 
which is at least one order of magnitude lower
than that in AGNs (\citealt{vV2016,Jiang2020}).
%To perform a fair comparison, w
We estimated the $f_d$ of turn-on AGNs
with the similar method used in \citet{Jiang2020}, which is $f_d=\ldust/\lbol$.
The \lbol\ of iPTF~16bco and SDSSJ1115+0544 after state transformation
are $\sim10^{45}$~\lum\ (\citealt{Gezari2017}) 
is $\sim4\times10^{44}$~\lum\ (\citealt{Yan2019}), respectively.
Meanwhile, their peak dust luminosity (\ldust) inferred from \wise\
W1 and W2 photometry (fitted by a blackbody radiation) 
is $\sim10^{44}$~\lum\ and $3\times10^{43}$~\lum, respectively.
%\shu{I am not sure whether the conclusion made below is strong enough, as 
%only two ``turn-on" AGNs are used.., please check and feel free to modify the 
%description..}
The inferred $f_d$ ($\sim0.1$) for "turn-on" AGNs appears to fall within the regime 
connecting normal galaxies and AGNs.  
The result is interesting, and may shed important light onto the triggering mechanism of AGNs,
which is perhaps regulated by the availability of interstellar medium in the vicinity of SMBHs. 
%Future observations of more ``turn-on'' AGNs will be helpful to test this connection. 

If "turn-on" AGNs occur only when a large amount of gas and dust accumulate 
around the SMBHs, the dust echo appears inevitable, explaining the high efficiency of 
searching in MIR.
Particularly in the extremely dusty environments, the ignited SMBHs
can be severely obscured and only be identified in the IR (\S\ref{obsTDE}).
Uncovering more \emph{bona fide} turn-on AGNs are valuable
not only for understanding the AGN accretion physics, but also
for the probe of AGN duty cycle as well as the impact on their host galaxies.
The duty cycle, namely the accreting phase of a SMBH (i.e., AGN phase) is
considered to be roughly a few times $10^7$~yr (\citealt{Combes2000,Haehnelt1993}).
However, the turn-on time scale is found within one year 
(\citealt{Gezari2017,Yan2019}),  indicating that the event rate could be 
as low as $\sim10^{-7}$.
Such a low expected rate seems contrary to the discovery of 
two and possibly more cases in our sample. 
Perhaps the traditional duty cycle of $10^7$~yr is only suitable for the total 
active time but not appropriate for flare events which may have different time scales.
The current and future surveys may be able to constrain the time scale distribution 
with a large number of newly discovered "turn-on" AGNs.
The specific mode of AGN duty cycle may have distinctive influence on 
their host galaxies and advance our understanding of galaxy evolution.

%\subsection{Characterizing the dust content in the vicinity of SMBHs}

\section{Summary and Prospects}

%summarize what we have done
The combined \wise\ and \neowise\ light curves, which have a time baseline of 
about one decade with a cadence of half year, have provided us 
a unique dataset for MIR time-domain study.  
Starting from  $\sim1$ million SDSS galaxies, we have selected out 137 galaxies which have 
displayed MIR outburst with amplitude $>0.5$ magnitude with respect 
to the pre-outburst quiescent phase.
Only a small fraction (15/137) of these outbursts have been reported 
by optical surveys including four SNe, two TDE candidates, three turn-on AGNs
and six unclassified objects.  
The remaining sources are likely associated with the dust echoes
of transient accretion events of SMBHs, as suggested by  
their proximity to host galaxy centers, their high MIR luminosity, 
weak radio emission and MIR luminosity distribution, {\it i.e.} luminosity function.
We are undertaking multi-wavelength follow-up observations to identify the 
nature of these MIR outburst in nearby galaxies (MIRONG).
The MIRONG unveiled by our study demonstrate the importance and necessity of MIR time-domain survey. 
For example, they may pave a way to solve the perplexed issues in the 
current study of TDEs and changing-look AGNs.

%Prospects in the future
Since the \neowise\ survey is on-going, we expect more data points will be 
accumulated in the future.
%as a conservative estimate. 
Additional data will help us 
diagnose the nature of MIR outbursts and obtain better measurements of dust properties. 
In addition, more MIR transients will be discovered with the updated database. 
%Looking ahead, 
At the post-WISE era, the Near-Earth Object Camera (NEOCam) 
is a planned mission to discover and characterize asteroids and comets at 
two MIR channels simultaneously, which will cover 68\% of the extragalactic sky at 
the wavelengths of $4.0-5.2~\mu$m and $6.0-10.0\mu$m, respectively (\citealt{Ross2019}). 
The survey depth of NEOCam is quite similar to \neowise\ yet its cadence is 
even better with on average 30 visits per year. 
Moreover, the already selected SPHEREx mission (\citealt{Dore2016}) 
that is scheduled to launch in 2023, will be an excellent complement to NEOCam at near-IR band, 
aiming to obtain spectra over $0.75-5\mu$m across the full sky.
It will scan the entire sky four times during its nominal 25-month mission life, though 
it is not as deep as \wise. Therefore, the prospects for the study of MIR transients 
are still very bright.
 \\

%\acknowledgments
We are grateful to the anonymous referee for his/her careful reading 
and nice comments, which have greatly improved the paper.
We thank Dr. Roc Cutri for many useful suggestions about the use of \wise\
and \neowise\ archival data all the time
and Dr. John Moustakas for his nice help for the usage of {\tt IDL/iSEDfit}.
This work is supported by the Chinese Science Foundation 
(NSFC-11833007, 12073025, 11421303, 11733001), 
Joint Research Fund in Astronomy (U1731104) under cooperative agreement
between the NSFC and the CAS, Anhui Provincial Natural Science Foundation 
and the Fundamental Research Funds for the Central Universities.
This research makes use of data products from the
\emph{Wide-field Infrared Survey Explorer}, which is a joint
project of the University of California, Los Angeles,
and the Jet Propulsion Laboratory/California Institute
of Technology, funded by the National Aeronautics and
Space Administration. This research also makes use of
data products from {\emph{NEOWISE-R}}, which is a project of the
Jet Propulsion Laboratory/California Institute of Technology,
funded by the Planetary Science Division of the
National Aeronautics and Space Administration.
This research has made use of the NASA/ IPAC
Infrared Science Archive, which is operated by the 
California Institute of Technology, under contract
with the National Aeronautics and Space Administration.
This research has made use of the NASA/IPAC Extragalactic Database (NED),
which is operated by the Jet Propulsion Laboratory, 
California Institute of Technology,
under contract with the National Aeronautics and Space Administration.

\clearpage

\clearpage

%\begin{longrotatetable}
\startlongtable
\begin{deluxetable*}{crrcccccccc}
\tablecaption{The properties of MIR flares}  
\label{tb1_sample} 
\tablewidth{0pt}
\tabletypesize{\scriptsize}
\tablehead{
\colhead{Name} & \colhead{RA}  & \colhead{DEC} & \colhead{$z$}  & \colhead{$\Delta$W1}  &  \colhead{$\Delta$W2}  & \colhead{W1m} & \colhead{W2m} & \colhead{$L_{\rm W1}$} & \colhead{$L_{\rm W2}$} & \colhead{$\Delta d$} \\
& & & & mag & mag & mag & mag & \lum\ & \lum\ & arcsec 
}
\colnumbers
\startdata
SDSSJ000046.46+143813.0 & 0.193583 & 14.6369 & 0.13660 & 0.32$\pm$0.05 & 0.58$\pm$0.05 & -24.19$\pm$0.14 & -25.37$\pm$0.08 & 43.19$\pm$0.06 & 43.27$\pm$0.03 & 0.08 \\
SDSSJ002701.03+071357.6 & 6.754291 & 7.23266 & 0.13109 & 0.37$\pm$0.03 & 0.60$\pm$0.06 & -23.91$\pm$0.07 & -24.94$\pm$0.10 & 43.08$\pm$0.03 & 43.10$\pm$0.04 & 0.02 \\
SDSSJ004500.47-004723.1 & 11.25195 & -0.7897 & 0.05677 & 0.72$\pm$0.04 & 1.17$\pm$0.16 & -21.99$\pm$0.06 & -22.93$\pm$0.10 & 42.31$\pm$0.02 & 42.30$\pm$0.04 & 0.43 \\
SDSSJ010320.42+140149.8 & 15.83508 & 14.0305 & 0.04181 & 1.29$\pm$0.03 & 1.90$\pm$0.02 & -25.34$\pm$0.01 & -26.30$\pm$0.01 & 43.65$\pm$0.01 & 43.64$\pm$0.00 & 0.07 \\
SDSSJ012047.99-082918.4 & 20.19995 & -8.4884 & 0.03468 & 0.42$\pm$0.03 & 0.55$\pm$0.02 & -24.10$\pm$0.07 & -25.02$\pm$0.04 & 43.16$\pm$0.03 & 43.13$\pm$0.02 & 0.07 \\
SDSSJ012100.67+140517.3 & 20.25279 & 14.0881 & 0.12938 & 0.36$\pm$0.03 & 0.61$\pm$0.07 & -23.35$\pm$0.12 & -24.88$\pm$0.14 & 42.85$\pm$0.05 & 43.08$\pm$0.06 & 0.11 \\
SDSSJ015804.75-005221.8 & 29.51979 & -0.8727 & 0.08044 & 2.16$\pm$0.04 & 2.71$\pm$0.11 & -24.60$\pm$0.02 & -25.56$\pm$0.03 & 43.35$\pm$0.01 & 43.35$\pm$0.01 & 0.09 \\
SDSSJ020552.16+000411.8 & 31.46733 & 0.06994 & 0.07649 & 1.17$\pm$0.01 & 1.76$\pm$0.09 & -24.81$\pm$0.02 & -25.85$\pm$0.03 & 43.44$\pm$0.01 & 43.46$\pm$0.01 & 0.09 \\
\enddata
\tablecomments{
\scriptsize
(1): SDSS name.
(2)-(4): RA, DEC and redshift given by SDSS.
(5)-(6): The amplitudes of variability in W1 and W2 bands.
(7)-(8): The peak absolute magnitude in W1 and W2.
(9)-(10): The peak monochromatic luminosity of W1 and W2.
(11): The offset distance between the maximum outburst epoch and quiescent state.}
\end{deluxetable*}
%\end{longrotatetable}

\clearpage

%\begin{longrotatetable}
\startlongtable
\begin{deluxetable*}{ccccccccc}
\tablecaption{The fitted dust parameters}  
\label{tb_dust} 
\tablewidth{0pt}
\tabletypesize{\scriptsize}
\tablehead{
\colhead{Name} & \colhead{$T_{\rm bb}$}  & \colhead{log$L_{\rm bb}$} & \colhead{log$M_{\rm bb}$}  & \colhead{log$R_{\rm bb}$}  & \colhead{$T_{\rm d}$}  &  \colhead{log$L_{\rm d}$}  & \colhead{log$M_{\rm d}$} & \colhead{log$R_{\rm d}$} \\
 & K & \lum\ & \msun\ & pc & K & \lum\ & \msun\ & pc   
}
\colnumbers
\startdata
SDSSJ0000+1438 &    901$\pm$40 & 43.41$\pm$0.09 & -2.01$\pm$0.12 & -1.12$\pm$0.09 &      794$\pm$31 & 43.37$\pm$0.09 & -1.07$\pm$0.11 & -0.65$\pm$0.12 \\
SDSSJ0027+0713 &    995$\pm$50 & 43.24$\pm$0.10 & -2.35$\pm$0.14 & -1.29$\pm$0.10 &      865$\pm$37 & 43.20$\pm$0.10 & -1.39$\pm$0.13 & -0.81$\pm$0.14 \\
SDSSJ0045-0047 &    996$\pm$129 & 42.45$\pm$0.25 & -3.14$\pm$0.34 & -1.69$\pm$0.24 &      857$\pm$93 & 42.42$\pm$0.25 & -2.20$\pm$0.31 & -1.22$\pm$0.33 \\
SDSSJ0103+1401 &    961$\pm$22 & 43.79$\pm$0.04 & -1.74$\pm$0.06 & -0.98$\pm$0.04 &      829$\pm$16 & 43.76$\pm$0.04 & -0.81$\pm$0.05 & -0.52$\pm$0.06 \\
SDSSJ0120-0829 &    998$\pm$27 & 43.29$\pm$0.05 & -2.31$\pm$0.07 & -1.27$\pm$0.05 &      855$\pm$20 & 43.25$\pm$0.04 & -1.37$\pm$0.06 & -0.80$\pm$0.07 \\
SDSSJ0121+1405 &    727$\pm$29 & 43.25$\pm$0.11 & -1.80$\pm$0.13 & -1.01$\pm$0.10 &      657$\pm$24 & 43.20$\pm$0.11 & -0.90$\pm$0.13 & -0.56$\pm$0.14 \\
SDSSJ0158-0052 &    999$\pm$93 & 43.50$\pm$0.18 & -2.10$\pm$0.25 & -1.16$\pm$0.17 &      862$\pm$68 & 43.46$\pm$0.18 & -1.15$\pm$0.23 & -0.68$\pm$0.24 \\
SDSSJ0205+0004 &    936$\pm$60 & 43.60$\pm$0.14 & -1.88$\pm$0.18 & -1.06$\pm$0.13 &      815$\pm$45 & 43.57$\pm$0.14 & -0.94$\pm$0.17 & -0.59$\pm$0.18 \\
\enddata
\tablecomments{
(1): Object name.
(2)-(4): Dust temperature, luminosity and mass fitted with blackbody model
at the epoch of W2 luminosity maximum. The dust mass is in unit of solar mass.
(5): The distance of the dust to the central heating source assuming 
spherically symmetric distribution and covering factor 1.
(6)-(9): similar to (2)-(5) but with the dust absorption coefficient considered.
}
\end{deluxetable*}

\begin{deluxetable*}{llllll}
\tabletypesize{\small}
\setlength{\tabcolsep}{0.04in}
\tablecaption{Optical counterparts}
\tablewidth{0pt}
\tablehead{
\colhead{Galaxy} & \colhead{Transient} & \colhead{Date} & \colhead{Offset} & \colhead{AT Designation} & \colhead{Classification}}
\colnumbers
\startdata
SDSSJ0045-0047 & PS16dsp       & 2016-08-03 & 0.39     & AT2016ezq  & ... \\  
SDSSJ0158-0052 & PS16dtm       & 2016-08-12 & 0.16     & SN2016ezh  & TDE \\
SDSSJ0841+0526 & ATLAS17nvk    & 2017-11-15 & $<0.5^a$ & AT2017jdg  & ... \\
SDSSJ0854+1113 & GNT           & 2016-09-26 & $<0.5^a$ & ...        & ...\\
SDSSJ0915+4814 & ZTF18aaidlyq  & 2018-11-03 & 0.13     & AT2018ivp  & Turn-on AGN \\
SDSSJ0936+0615 & ASASSN-18xl   & 2018-10-09 & 2.93     & SN2018hfm  & SNe \\
%SDSSJ1008+1504 & PS15mv        & 2015-03-14 & 0.20     & ...        & ... \\
SDSSJ1133+6701 & ZTF18aasuray  & 2018-05-10 & 0.09     & AT2018cdp  & Turn-on AGN \\
%SDSSJ1328+6227 & ZTF18aakdyss  & 2018-04-16 & 0.12     & AT2018bjj  &  ... \\
SDSSJ1332+2036 & CRTS\_OBC     & 2015-03-29 & ...      & ...        & ... \\
\enddata
\tablecomments{
(1): SDSS abbreviated Name of the galaxy.
(2): The transient named by the discovering surveys.
(3): Optical discovery date.
(4): Distance to the galactic center in unit of arcsec.
(5): Designation on the transient name server.
(6): The classified nature of the transients. 
}
\tablenotetext{a}{The precise offset is unknown but should be $<0\arcsec.5$ because of the selection criteria of Gaia nuclear transient (Kostrzewa-Rutkowska et al. 2018)}
\label{optical_match}
\end{deluxetable*}

\clearpage

\startlongtable
\begin{deluxetable*}{crrccccl}
\tablecaption{The properties of host galaxies and black hole}
\label{tb2_host}
\tablewidth{700pt}
%\tabletypesize{\scriptsize}
\tablehead{
\colhead{Name} & \colhead{$M_{r}$}  & \colhead{g-r} & \colhead{$\sigma_{\star}$} & \colhead{log$M_{*}$} & \colhead{logSFR} &  \colhead{log\mbh} & \colhead{Type} 
}
\colnumbers
\startdata
SDSSJ0000+1438 & -21.54 & 0.89 & 192. & 11.11 & -0.13 & $8.23$ & Seyfert \\
SDSSJ0027+0713 & -21.48 & 0.96 & 154. & 11.05 & 0.89 & $7.68$ & Seyfert \\
SDSSJ0045-0047 & -19.48 & 0.41 & 44.9 & 9.604 & -0.10 & $6.00$ & StarForming \\
SDSSJ0103+1401 & -21.39 & 0.71 & 94.2 & 10.73 & 0.94 & $6.47$ & StarForming \\
SDSSJ0120-0829 & -20.83 & 0.79 & 115. & 10.73 & 0.80 & $6.79$ & Seyfert(b) \\
SDSSJ0121+1405 & -21.49 & 0.77 & 130. & 10.91 & 0.49 & $7.26$ & StarForming \\
SDSSJ0158-0052 & -19.49 & 0.61 & 45.4 & 10.15 & -0.09 & $6.37$ & StarForming(b) \\
SDSSJ0205+0004 & -20.44 & 0.71 & 112. & 10.55 & 0.56 & $6.92$ & StarForming \\
\enddata
\tablecomments{
(1): SDSS name.
(2)-(3): SDSS $r$-band absolute magnitude and $g-r$ color after $k$-correction to 0.1 .
(4): Stellar velocity dispersion.
(5): Stellar mass.
(6): Estimated BH mass.
(7): Classification in the BPT-diagram with suffix '(b)' highlighting robust broad \ha\ detection.
(8): The logarithmic star formation rate in unit of $\msun/\rm yr$ (see \S\ref{Jet}).}
\end{deluxetable*}

\begin{deluxetable*}{clrrrrlcl}
\tabletypesize{\small}
\setlength{\tabcolsep}{0.06in}
\tablecaption{Radio properties of FIRST-detected objects}
\tablewidth{0pt}
\tablehead{
Name & Type & $F_{\rm 1.4}$ & log$L_{\rm 1.4}$ & logSFR & $F^{'}_{\rm 1.4}$ & $F^{'}_{\rm 1.4}$/$F_{\rm 1.4}$ & $R$ & Class}
\colnumbers 
\startdata
SDSSJ0120-0829 & Seyfert(b) & 3.68 &    22.00 &   0.80 &   8.00 &   0.46 &     12 & SF \\
SDSSJ0205+0004 & StarForming & 0.75 &    22.02 &   0.56 &   0.76 &   0.98 & ... & SF \\
SDSSJ0745+2655 & Seyfert(b) & 3.09 &    23.01 &   1.60 &   7.97 &   0.39 &      3 & SF \\
SDSSJ0811+4054 & Seyfert(b) & 1.64 &    22.24 &   1.50 &  18.02 &   0.09 &     10 & SF \\
SDSSJ1046+1655 & LINER & 69.72 &    24.91 &   0.97 &   0.32 & 216 &  29226 & Jet \\
SDSSJ1105+5941 & Seyfert(b) & 5.96 &    22.19 &   0.66 &   5.62 &   1.06 &     12 & w.Jet \\
SDSSJ1114+4056 & Seyfert(b) & 1.28 &    22.89 &   0.66 &   0.24 &   5.30 &     26 & w.Jet \\
SDSSJ1129+5131 & Composite & 5.01 &    22.09 &   0.45 &   3.06 &   1.64 & ... & SF \\
\enddata
\tablecomments{
(1): Name.
(2): BPT type listed in Table~\ref{tb2_host}.
(3): Radio flux at 1.4~GHz in unit of mJy.
(4): The logarithmic radio luminosity at 1.4~GHz in unit of $\rm W~Hz^{-1}$
after k-correction.
(5): SFR estimated from SDSS+\wise\ SED fitting or W4 flux. 
(6): Radio flux derived from SFR.
(7): The ratio between the observed radio flux and the predicted one from star formation.
(8): The radio-loudness parameter defined as the ratio of the 
monochromatic flux between 6~cm (5~GHz) and AGN continuum at 4400~\AA\ 
($f_{\nu,\rm 6cm}/f_{\nu,\rm 4400\AA}$). 
(9): The main origin of radio emission. 'SF': star formation. 'Jet': undisputed intense 
jet emission. 'w.Jet': jet likely contribute since the observed radio flux is higher than
SFR prediction, but the radio power should be much weaker than 'Jet' as $R$ is not very high
(<30).
}
\label{radio}
\end{deluxetable*}

\end{document}